\documentclass[journal,twoside,web]{ieeecolor}
\usepackage{generic}
\usepackage{cite}
\usepackage{amsmath,amssymb,amsfonts}
\usepackage{algorithmic}
\usepackage{graphicx}
\usepackage{algorithm,algorithmic}
\let\labelindent\relax
\usepackage{enumitem}
\usepackage{mathdots}
\usepackage{xfrac}
\usepackage{array}
\usepackage{balance} 
\usepackage{hyperref}
\hypersetup{hidelinks=true}
\usepackage{textcomp}
\usepackage{bm}
\usepackage{bbm}
                        
\usepackage{color}

\newcommand{\hop}{\mathsf{H}}
\newtheorem{assumption}{Assumption}
\newtheorem{lemma}{Lemma}
\newtheorem{remark}{Remark}
\newtheorem{definition}{Definition}
\newtheorem{corollary}{Corollary}

\newtheorem{theorem}{Theorem}

\usepackage{eso-pic} 
\AddToShipoutPictureBG*{% 
\AtPageUpperLeft{% 
\setlength\unitlength{1in}% 
\hspace*{\dimexpr0.5\paperwidth\relax} 
\makebox(0,-0.75)[c]{ 
\begin{tabular}{c c} 
Rodrigo A. Gonz\'alez \emph{et al.},

Finite Sample MIMO System Identification with Multisine Excitation: Nonparametric, Direct, \\ and Two-step Parametric Estimators, 

{\em submitted for journal publication}, 
uploaded to ArXiv \today \\ 
\end{tabular}}}}

\def\BibTeX{{\rm B\kern-.05em{\sc i\kern-.025em b}\kern-.08em
    T\kern-.1667em\lower.7ex\hbox{E}\kern-.125emX}}
\begin{document}

\title{Finite Sample MIMO System Identification with Multisine Excitation: Nonparametric, Direct, and Two-step Parametric Estimators}
\author{Rodrigo A. Gonz\'alez, \IEEEmembership{Member, IEEE}, Koen Classens, \IEEEmembership{Graduate Student Member, IEEE}, Cristian R. Rojas, \IEEEmembership{Senior Member, IEEE}, Tom Oomen, \IEEEmembership{Senior Member, IEEE}, H\aa kan Hjalmarsson, \IEEEmembership{Fellow, IEEE}
\thanks{R. A. Gonz\'alez, K. Classens, and T. Oomen are with the Department of Mechanical Engineering, Eindhoven University of Technology, Eindhoven, The Netherlands. T. Oomen is also with the Delft Center for Systems and Control, Delft University of Technology, Delft, The Netherlands (e-mails: r.a.gonzalez@tue.nl; k.h.j.classens@tue.nl; t.a.e.oomen@tue.nl). }
\thanks{C. R. Rojas and H. Hjalmarsson are with Division of Decision and Control Systems, KTH Royal Institute of Technology, Stockholm, Sweden (e-mails: crro@kth.se; hjalmars@kth.se).}}
\maketitle
\begin{abstract}
Multisine excitations are widely used for identifying multi-input multi-output systems due to their periodicity, data compression properties, and control over the input spectrum. Despite their popularity, the finite sample statistical properties of frequency-domain estimators under multisine excitation, for both nonparametric and parametric settings, remain insufficiently understood. This paper develops a finite-sample statistical framework for least-squares estimation of the frequency response function (FRF) and its implications for parametric modeling. First, we derive exact distributional and covariance properties of the FRF estimator, explicitly accounting for aliasing effects under slow sampling regimes, and establish conditions for unbiasedness, uncorrelatedness, and consistency across multiple experiments. Second, we show that the FRF estimate is a sufficient statistic for any parametric model under Gaussian noise, leading to an exact equivalence between optimal two stage frequency-domain methods and time-domain prediction error and maximum likelihood estimation. This equivalence is shown to yield finite-sample concentration bounds for parametric maximum likelihood estimators, enabling rigorous uncertainty quantification, and closed-form prediction error method estimators without iterative optimization. The theoretical results are demonstrated in a representative case study.
\end{abstract}

\section{Introduction}
\label{sec:introduction}
\IEEEPARstart{I}{n} system identification, multisine input excitation offers significant advantages over white noise excitation for retrieving informative output data of an unknown system. These advantages include data compression, the simplification of the model validation step, and full control of the input power spectrum \cite{schoukens1994advantages}. Furthermore, by exploiting the periodicity of multisine input signals, time-averaging and multiple experiments can be used to reduce the variance of the Discrete Fourier transform of the output, hereby improving the signal-to-noise ratio~\cite{schoukens2004time}.

Multisines are closely linked to frequency-domain identification approaches \cite{pintelon2012system}, since they enable the direct estimation of the frequency response function (FRF) of the system at each input frequency due to their sparse spectrum. From FRF estimates, it is common to identify low-order parametric models through fitting the frequency-domain data. Levy \cite{levy1959complex} proposed a weighted least-squares approach that avoids non-convex cost functions to be minimized, Sanathanan and Koerner (SK) developed a method based on least-squares iterations \cite{sanathanan1963transfer}, and instrumental variable iterations have been studied in \cite{van2014optimally}. Asymptotic analyses of iterative least-squares methods were pursued in \cite{whitfield1987asymptotic}, and vector fitting methods have been developed in \cite{gustavsen1999rational} and \cite{ozdemir2017transfer}, which improve on the numerical conditioning of the SK iterations by using barycentric basis functions, and were compared in \cite{voorhoeve2014numerically}. Extensions to multi-input multi-output (MIMO) systems are also available in the literature \cite{de1996multivariable,gaikwad1997multivariable,blom2010multivariable,voorhoeve2016estimating}, with applications to, e.g., modal identification of civil aircraft structures \cite{vayssettes2014frequency} and high-tech mechatronic systems \cite{voorhoeve2020identifying}. Most of these methods for MIMO systems incorporate weighting functions acting elementwise on the multivariate error, known as Schur weighting \cite{de1996multivariable}. While this approach is intuitive and permits focusing the estimation on the bandwidth of interest, ad-hoc weights may compromise the statistical properties of the resulting parametric model.

Despite the wide use of frequency-domain system identification methods, systematic studies of the finite-sample statistical properties of FRF estimation and their implications for parametric modeling remain scarce. Most analyses have been conducted in the asymptotic regime \cite{antoni2007comprehensive,pintelon2010estimation,pintelon2011improved}, or only consider statistics of single-input single-output FRF measurements in errors-in-variables settings \cite{pintelon2001measurement,pintelon2002probability}. Earlier surveys on iterative methods, as well as maximum-likelihood and least-squares estimators with frequency domain data, were put forward in \cite{pintelon1994parametric} and \cite{mckelvey2002frequency} respectively, while relations between frequency-domain and time-domain identification methods have been covered in \cite{schoukens1999study,aguero2010equivalence}, with statistical foundations being presented in \cite{brillinger2001time}. In contrast, finite sample results in other identification contexts have risen prominently in recent years \cite{simchowitz2018learning,gonzalez2020deviation,jedra2022finite}. In \cite{hjalmarsson2006least}, the variance of the least-squares estimator for scalar frequency response functions for discrete-time systems was computed explicitly. Our results, while related to \cite{hjalmarsson2006least}, focus on MIMO continuous-time systems, incorporate the effect of undersampling and multiple experiments, and provide a finite-sample link between nonparametric and parametric estimators when the input excitation is a continuous-time multisine signal. 

Recent studies \cite{tsiamis2024finite,tsiamis2024finiteb} have derived finite-sample convergence rates for the empirical transfer function estimate (ETFE, \cite{ljung1998system}), and related them to error rate orders for the resulting impulse response estimate. Their approach differs fundamentally from this paper in several ways. First, we do not explicitly address the ETFE, but a least-squares estimator of the band-limited equivalent discrete-time impulse response \cite{gonzalez2021noncausal}, from which we derive finite-sample confidence bounds for parametric prediction error and maximum likelihood estimators, and establish novel formal relationships between time-domain and frequency-domain parametric system identification. We also consider continuous-time multisines, do not require a fixed frequency resolution, and explicitly address aliasing effects produced by undersampling \cite{gonzalez2024sampling}, which are relevant in applications such as vision-in-the-loop systems~\cite{van2023beyond}.

In summary, the main results of this paper are:
\begin{enumerate}[label=(C\arabic*)]
\label{contribution1}
	\item 
    We derive explicit finite-sample statistical properties of least-squares estimators for the FRF of a MIMO continuous-time system, including the exact distribution and the effect of aliasing under slow sampling. By relating the system to a band-limited discrete-time equivalent model, we determine the well-posedness and unbiasedness (Theorem \ref{thm2}), as well as the covariance structure and uncorrelateness of the estimates (Theorem~\ref{thm3}).
\label{contribution2}
    \item We establish conditions for weak consistency of the FRF estimate at the input frequencies as the number of samples or experiments grows (Corollary \ref{coroconsistency}), and provide a frequency-domain interpretation of the least-squares estimate (Theorem \ref{thmfrequency}).
    
    \item We prove that the FRF estimate is a sufficient statistic for any parametric model under Gaussian noise (Theorem \ref{thm4}), implying an exact equivalence between optimal two step frequency domain methods and time domain prediction error or maximum likelihood estimation for finite data (Corollary \ref{cor32}).

    \item We obtain conditions for which the maximum likelihood estimator of a parametric model can be computed in closed-form (Corollary \ref{cor33}), and determine finite-sample concentration bounds for the parametric FRF estimator, as well as for the parameter vector itself (Theorem \ref{thmprobability}). These results, which enable rigorous uncertainty quantification, lead to explicit formulas for the prediction error method in left matrix fraction description models (Theorem~\ref{thm6}), and are exemplified via a case study (Section~\ref{subsec:finitetime}).
\end{enumerate}

The remainder of this work is organized as follows. In Section \ref{sec:problemformulation} we present the system setup and identification setting. The least-squares estimator of MIMO frequency response functions is derived and analyzed in Section \ref{sec:ls}, while Section \ref{sec:parametric} uses these results to formulate and analyze statistically optimal parametric estimation methods. Section \ref{sec:parametrizations} contains examples on specific model parametrizations and case studies, while final remarks are provided in Section \ref{sec:conclusions}. Technical lemmas and their proofs can be found in the Appendix.

\textit{Notation}: The Heaviside operator $p$ satisfies $p\mathbf{x}(t)=\textnormal{d}\mathbf{x}(t)/\textnormal{d}t$. The Laplace transform of a signal $\mathbf{x}(t)$ is denoted $\mathcal{L}\{\mathbf{x}(t)\}(s)$, and it is a function of the complex variable $s$. All vectors, matrices, and multivariate signals are written in bold, and vectors are column vectors, unless transposed. The complex conjugate and conjugate transpose of a vector $\mathbf{x}$ is denoted as $\overline{\mathbf{x}}$ and $\mathbf{x}^\hop$ respectively, $\mathbf{A}^\dagger$ denotes the Moore-Penrose inverse of the matrix $\mathbf{A}$, and the block diagonal matrix with matrices $\mathbf{A}_1,\dots,\mathbf{A}_m$ along its diagonal is denoted as $\textnormal{blkdiag}(\mathbf{A}_1,\dots,\mathbf{A}_m)$. The operators $\otimes$ and $\odot$ denote the Kronecker and element-wise products, respectively, the 2-norm is denoted by $\|\mathbf{x}\|$, the weighted 2-norm by $\|\mathbf{x}\|_{\mathbf{A}}=\sqrt{\mathbf{x}^\top \mathbf{A}\mathbf{x}}$, and the Frobenius norm by $\|\mathbf{A}\|_{\textnormal{F}}$. The vector formed by $m$ ones is denoted $\mathbf{1}_m$, and the identity matrix of size $m$ is denoted $\mathbf{I}_m$. The minimum singular value of $\mathbf{A}$ is $\sigma_{\textnormal{min}}(\mathbf{A})$, and if $\mathbf{A}$ is Hermitian, its minimum eigenvalue is denoted as $\lambda_{\textnormal{min}}(\mathbf{A})$. The elementwise vector-to-vector cosine and exponential functions are denoted $\cos(\mathbf{x})$ and $e^{\mathbf{x}}$ respectively. The floor function is denoted as $\lfloor x\rfloor$, and the Kronecker delta function is denoted as $\delta_K(n)$, with $n\in\mathbb{Z}$.

\section{System and model setup}
\label{sec:problemformulation}
Consider the following MIMO, linear time-invariant, asymptotically stable, causal, continuous-time system
\begin{equation}
\label{system}
    \mathbf{x}(t) = \mathbf{G}_0(p)\mathbf{u}(t) ,
\end{equation}
where $\mathbf{x}(t)\in\mathbb{R}^{n_y}$ is the output of the system $\mathbf{G}_0(p)$ when subject to the input $\mathbf{u}(t)\in\mathbb{R}^{n_u}$. Our interest is centered in estimating $\mathbf{G}_0(p)$ from input and output data. Thus, we perform $m\geq n_u$ experiments with continuous-time multisine input excitations $\mathbf{u}_i(t)$ of the form
\begin{equation}
\label{input}
    \mathbf{u}_i(t)\hspace{-0.03cm}=\hspace{-0.03cm}\mathbf{a}_{0,i} + \sum_{\ell =1}^M \mathbf{a}_{\ell,i}\hspace{0.02cm}\odot\hspace{0.02cm} \cos(\omega_\ell \mathbf{1}_{n_u} t + \bm{\phi}_{\ell,i}), \quad \hspace{-0.03cm} i\hspace{-0.03cm}=\hspace{-0.03cm}1,\dots,m,
\end{equation}
where the frequencies $\omega_\ell$ are assumed positive and sorted in strictly ascending order, and $\mathbf{a}_{0,i}, \mathbf{a}_{\ell,i},\bm{\phi}_{\ell,i} \in \mathbb{R}^{n_u}$ for $\ell=1,2,\dots,M$, and $i=1,\dots,m$. Note that the input for each experiment is allowed to differ in the amplitudes $\mathbf{a}_{\ell,i}$ and in the phase components $\bm{\phi}_{\ell,i}$, but not in the angular frequencies $\omega_\ell$. The input coefficients are assumed to satisfy the following condition, which is commonly met in practical applications~\cite{pintelon2012system}:
\begin{assumption}[Input linear independence]
\label{assumption1}
    The following matrices have full column rank:
\begin{equation}
        \label{amatrices}
    \mathbf{A}_0 \hspace{-0.05cm}:= \hspace{-0.07cm}\begin{bmatrix}
	\mathbf{a}_{0,1}^\top \\
	\mathbf{a}_{0,2}^\top \\
	\vdots \\
	\mathbf{a}_{0,m}^\top
 \end{bmatrix}\hspace{-0.03cm}, \hspace{0.05cm}  \mathbf{A}_\ell\hspace{-0.04cm}:  \hspace{-0.02cm}= \hspace{-0.08cm} \begin{bmatrix}
	\frac{\mathbf{a}_{\ell,1}^\top}{2}  \\
	\frac{\mathbf{a}_{\ell,2}^\top}{2}  \\
	\vdots \\
	\frac{\mathbf{a}_{\ell,m}^\top}{2}
 \end{bmatrix}\odot \begin{bmatrix}
	e^{\mathrm{i}\bm{\phi}_{\ell,1}^\top}  \\
	e^{\mathrm{i}\bm{\phi}_{\ell,2}^\top}  \\
	\vdots \\
	e^{\mathrm{i}\bm{\phi}_{\ell,m}^\top}
 \end{bmatrix}\hspace{-0.03cm}, \hspace{0.03cm} \ell\hspace{-0.02cm}=\hspace{-0.02cm}1,\dots,M. 
\end{equation}
\end{assumption}
The following three conditions, when jointly satisfied, are sufficient to ensure that Assumption \ref{assumption1} holds with probability~1: (1) a subset of $n_u$ input offset vectors $\{\mathbf{a}_{0,i}\}_{i=1}^m$ forms a linearly independent set, (2) each element of each $\mathbf{a}_{\ell,i}$ is non-zero, and (3) the phases are chosen uniformly at random. 

For each experiment $i$, we retrieve a noisy measurement of the output vector $\mathbf{x}_i(t)$ at times $t=h,2h,\dots,Nh$:
\begin{equation}
\label{output}
    \mathbf{y}_i(kh)=\mathbf{x}_i(kh)+\mathbf{v}_i(kh), \quad k=1,\dots,N,
\end{equation}
where $\mathbf{v}_i(kh)\in\mathbb{R}^{n_y}$ is zero-mean i.i.d. white noise that has a positive definite covariance matrix $\bm{\Sigma}$, and where we assume that the noiseless output $\mathbf{x}_i(kh)$ is in a \textit{stationary regime}, i.e., no distortions associated with transient effects are considered. Importantly, the sampling period $h$ is not restricted to satisfy the Nyquist–Shannon criterion $h<\pi/\omega_M$, meaning that the measured output may suffer from aliasing effects \cite{oppenheim1997signals}. In cases where this condition is violated, we adopt an additional assumption under slow sampling to ensure that the input frequency lines do not overlap after aliasing. This guarantees a unique characterization of the input intersample behavior given the input samples and input frequency knowledge.

\begin{assumption}[No overlapping input frequencies]
\label{assumption2}
The input frequencies $\{\omega_\ell\}_{\ell=1}^M$ satisfy
    \begin{equation}
    \label{frequencycondition}
           \hspace{-0.4cm}\begin{cases}
               \omega_\ell\pm \omega_\tau \hspace{-0.06cm}\neq \hspace{-0.06cm} \frac{2 n \pi}{h} &\hspace{-0.22cm}\textnormal{for all } \ell,\tau\hspace{-0.04cm}=\hspace{-0.04cm} 1,\dots, M; \ell\neq \tau; n\in \mathbb{Z}, \\
               \omega_\ell\neq  \frac{n \pi}{h} &\hspace{-0.22cm}\textnormal{for all } \ell = 1,\dots, M; n\in \mathbb{Z}.
           \end{cases}
           \hspace{-0.4cm}
    \end{equation}
\end{assumption}
\begin{figure}
	\centering{
		\includegraphics[width=0.48\textwidth]{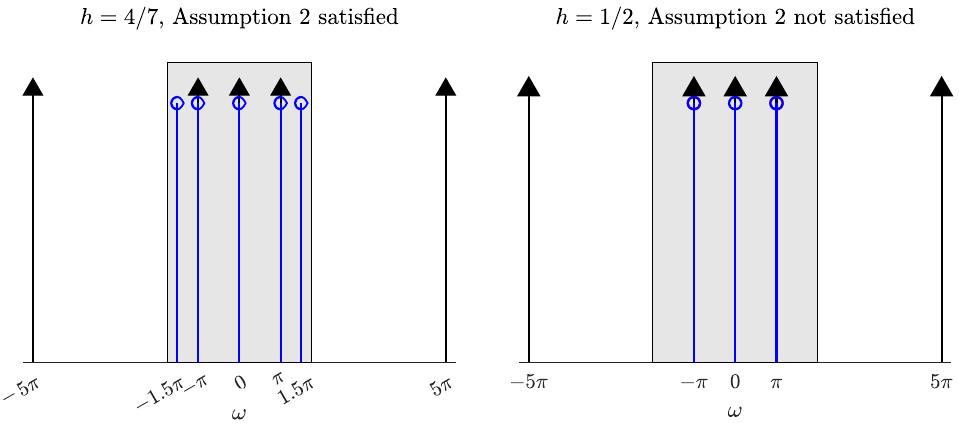}
        \vspace{-0.2cm}
		\caption{Input spectrum plots with sampling periods $h=4/7$ (left) and $h=1/2$ (right), representing scenarios where Assumption \ref{assumption2} is satisfied and violated, respectively. The black arrows indicate the magnitude spectrum of a continuous-time multisine input with frequencies $\omega_\ell =  0, \pi$, and $5\pi$. The blue stem plots show the discrete-time spectrum, and the grey boxes denote the fundamental frequency band.}
		\label{fig_nyquist}}
        \vspace{-0.3cm}
\end{figure}
Figure \ref{fig_nyquist} illustrates examples where Assumption \ref{assumption2} is either satisfied or violated. Intuitively, Assumption \ref{assumption2} ensures that each input frequency maps to a unique frequency in the discrete-time fundamental band, preserving the number of spectral lines after sampling. Interestingly, a larger sampling period does not imply the violation of the condition in Assumption \ref{assumption2}.

Our goal is to derive and analyze estimators for $\mathbf{G}_0(p)$ given samples of the continuous-time input signals $\mathbf{u}_i(t)$ and their corresponding noisy output samples $\{\mathbf{y}_i(kh)\}_{k=1}^{N}$ for $i=1,\dots,m$. We will examine three estimators and their relationships: a nonparametric estimator of the FRF $\mathbf{G}_0(\mathrm{i}\omega)$, a parametric estimator $\mathbf{G}(p,\bm{\theta})$ (parameterized by a vector $\bm{\theta}\in \mathbb{R}^{n_\theta}$) derived directly from the time domain data, and a second parametric estimator which constructs a model of $\mathbf{G}_0(p)$ based solely on the nonparametric estimator of the FRF.

\begin{remark}
    The marginally stable system case is not explicitly addressed in this paper, though similar results to Theorems~\ref{thm2} to \ref{thm4} can be derived for this scenario. In this case, the input should not contain frequency component matching the poles of $\mathbf{G}_0(p)$ in the imaginary axis, and achieving output stationarity in practice may be more challenging due to resonant behavior.
\end{remark}
\begin{remark}
    For all the results in this paper, the single-input multi-output (SIMO) and single-input single-output (SISO) cases are recovered as a special case by setting $n_u=1$ and $n_u=n_y=1$, respectively. For these cases only one identification experiment (i.e., $m=1$) is required. Nonetheless, we also accommodate the use of multiple experiments, in line with the frameworks proposed in \cite{markovsky2015identification,vasquez2022consistency} for discrete-time parametric system identification.
\end{remark}

\section{Least-squares estimator of the MIMO frequency response function}
\label{sec:ls}
This section introduces a multisine equivalent discrete-time form of the system \eqref{system}, which is later used for system identification. Unlike zero-order hold \cite{aastrom1984computer} and band-limited discrete-time equivalents \cite{gonzalez2021noncausal}, the multisine equivalent discrete-time model is always finite-dimensional, and it depends on the input frequencies $\omega=0,\omega_1,\dots,\omega_M$. It accurately incorporates a multisine intersample behavior without reconstruction error, providing exact output values at the sampling instants $t=kh$, $k\in\mathbb{N}$. We compute the least-squares estimator of this discrete-time equivalent system and determine its finite-sample and asymptotic statistical properties in Section \ref{sub:statisticalpropertiesnonparametric}, followed by a frequency-domain interpretation in Section~\ref{sub:frequency}. 

\subsection{Multisine discrete-time equivalent system}

We introduce the multisine equivalent discrete-time response and present its main properties.
\begin{definition}
\label{definitionmultisine}
Given a sampling period $h$ and input frequencies $\omega=0,\omega_1,\dots,\omega_M$, we define the multisine equivalent discrete-time response of a causal MIMO system with continuous-time impulse response $\mathbf{H}(t)=\mathcal{L}^{-1}\{\mathbf{G}(s)\}(t)$ as
	\begin{equation}
	\label{hms}
	\mathbf{H}_{\textnormal{MS}}(kh) \hspace{-0.04cm}:= \hspace{-0.04cm}\int_{0}^{\infty}\hspace{-0.1cm}\eta_{k}(\tau)\mathbf{H}(\hspace{-0.01cm}\tau\hspace{-0.01cm})  \textnormal{d}\tau, \hspace{-0.07cm}\quad k=0,1,\dots,2M,
	\end{equation}
where $\eta_{k}(\tau)\in \mathbb{R}$ is the $(k+1)$-th entry of the vector solving the interpolation problem $\bm{\eta}^\top(\tau)\tilde{\bm{\Gamma}} = \bm{\varphi}^\top(\tau)$, where 
\begin{align}
\label{varphi}
&\bm{\varphi}(\tau) = [1, e^{-\mathrm{i}\omega_1 \tau},e^{\mathrm{i}\omega_1 \tau}, \dots,  e^{-\mathrm{i}\omega_M \tau},e^{\mathrm{i}\omega_M \tau}]^\top, \\
\label{gammatilde}
\tilde{\bm{\Gamma}} \hspace{-0.06cm}&=\hspace{-0.1cm} \begin{bmatrix}
1\hspace{-0.05cm} & \hspace{-0.05cm}1 & 1 & \hspace{-0.05cm}\cdots\hspace{-0.05cm} & 1 & 1 \\
1\hspace{-0.05cm} & \hspace{-0.05cm}e^{-\mathrm{i}\omega_1 h} & e^{\mathrm{i}\omega_1 h} & \hspace{-0.05cm}\cdots\hspace{-0.05cm} & e^{-\mathrm{i}\omega_M h} & e^{\mathrm{i}\omega_M h} \\
\vdots\hspace{-0.05cm} & \hspace{-0.05cm}\vdots & \vdots & & \vdots & \vdots \\
1\hspace{-0.05cm} & \hspace{-0.05cm}e^{-2\mathrm{i}\omega_1 Mh} & e^{2\mathrm{i}\omega_1 Mh} & \hspace{-0.05cm}\cdots\hspace{-0.05cm} & e^{-2\mathrm{i}\omega_M M h} & e^{2\mathrm{i}\omega_M Mh}
\end{bmatrix}\hspace{-0.06cm}.
\end{align}

\end{definition}
\begin{remark}
The multisine equivalent discrete-time response is well defined if and only if the matrix $\tilde{\bm{\Gamma}}$ in \eqref{gammatilde} is nonsingular. This matrix is a Vandermonde matrix, known to have full rank if the complex numbers $1,e^{\pm \mathrm{i}\omega_1 h},\dots, e^{\pm \mathrm{i}\omega_M h}$ are distinct \cite[Sec. 0.9.11]{Horn2012}. Such condition is equivalent to \eqref{frequencycondition} in  Assumption~\ref{assumption2}, which is satisfied irrespectively of $\omega_1,\dots,\omega_{M-1}$ if the Nyquist-Shannon condition $h<\pi/\omega_M$ is met. Additionally, the entries of the vector $\bm{\eta}(\tau)$ are multisines of frequencies $\omega=0,\omega_1,\dots,\omega_M$ and are real-valued for any $\tau\in \mathbb{R}$, since the conjugate vector $\overline{\bm{\eta}(\tau)}$ satisfies $\bm{\eta}^\hop(\tau)\tilde{\bm{\Gamma}} = \bm{\varphi}^\top(\tau)$.
\end{remark}

The main properties of the discrete-time equivalent response introduced in Definition \ref{definitionmultisine} require expressing the multisine input in terms of prior input samples. To this end, we represent the time-shifted input of the $i$th experiment in its complex exponential form
\begin{align}
\label{uitau}
\mathbf{u}_i(t-\tau) =(\bm{\varphi}^\top(\tau) \otimes \mathbf{I}_{n_u})\bm{\zeta}_i(t),
\end{align}
where $t,\tau\in \mathbb{R}$, $\bm{\varphi}(\tau)$ is defined in \eqref{varphi}, and
\begin{equation}
\label{zeta}
    \bm{\zeta}_i(t) =  \begin{bmatrix}
        \mathbf{a}_{0,i} \\
        \frac{\mathbf{a}_{1,i}}{2}\odot e^{\mathrm{i}(\omega_1 \mathbf{1}_{n_u} t+\bm{\phi}_{1,i})} \\
        \frac{\mathbf{a}_{1,i}}{2}\odot e^{-\mathrm{i}(\omega_1 \mathbf{1}_{n_u} t+\bm{\phi}_{1,i})} \\
        \vdots \\
        \frac{\mathbf{a}_{M,i}}{2}\odot e^{\mathrm{i}(\omega_M \mathbf{1}_{n_u}  t+\bm{\phi}_{M,i})} \\
        \frac{\mathbf{a}_{M,i}}{2}\odot e^{-\mathrm{i}(\omega_M \mathbf{1}_{n_u}  t+\bm{\phi}_{M,i})}
    \end{bmatrix}.
\end{equation}
By exploiting the fact that $\tilde{\bm{\Gamma}}$ in \eqref{gammatilde} is a matrix whose rows are given by $\bm{\varphi}^\top(0),\bm{\varphi}^\top(h),\dots,\bm{\varphi}^\top(2Mh)$, the vector $\bm{\zeta}_i(t)$ is a linear combination of $2M+1$ input samples via the relation
\begin{equation}
\label{bigui2}
    \bm{\zeta}_i(t) = (\tilde{\bm{\Gamma}}^{-1}\otimes \mathbf{I}_{n_u}) \mathbf{U}_i(t),
\end{equation}
where $\mathbf{U}_i(t)=[\mathbf{u}_i^\top(t),\mathbf{u}_i^\top(t-h), \dots, \mathbf{u}_i^\top(t-2Mh)]^\top$.

Lemma \ref{lem1} presents the link between continuous-time and discrete-time convolution with the multisine equivalent response, and introduces the discrete-time finite impulse response (FIR) model that characterizes the continuous-time system for multisine excitations.
\begin{lemma}
	\label{lem1}	
	Assume that Assumption \ref{assumption2} holds, and consider the discrete-time transfer function
	\begin{equation}
 \label{definitionGMS}
	\mathbf{G}_{\textnormal{MS}}(z) := \sum_{k=0}^{2M} \mathbf{H}_{\textnormal{MS}}(kh)z^{-k},
	\end{equation}
	with $\mathbf{H}_{\textnormal{MS}}(kh)$ being defined as in \eqref{hms}. Then, the following statements are true:
 \begin{enumerate}
        \item For any multisine input $\mathbf{u}_i(t)$ of the form in \eqref{input}, the output of $\mathbf{G}(p)$ satisfies
        \begin{align}
        \label{convolutionct}
            \mathbf{x}_i(t) &= \int_0^\infty \mathbf{H}(\tau)\mathbf{u}_i(t-\tau)\textnormal{d}\tau  \\
            \label{convolutionms}
            &= \sum_{r=0}^{2M} \mathbf{H}_{\textnormal{MS}}(rh)\mathbf{u}_i(t-rh).
        \end{align}
        \item The frequency response function $\mathbf{G}(p)$ of the continuous-time system with impulse response $\mathbf{H}(t)$ satisfies, for $\ell =1,\dots, M$,
	\begin{equation}
	\label{ch9:frequencyresponse}
	\hspace{-0.3cm}\mathbf{G}(0) = \mathbf{G}_{\textnormal{MS}}(1),\hspace{0.1cm} \mathbf{G}(\pm\mathrm{i}\omega_{\ell}) = \mathbf{G}_{\textnormal{MS}}(e^{\pm\mathrm{i}\omega_\ell h}).
	\end{equation}
\end{enumerate}
\end{lemma}

\textit{Proof of (1).} By inserting \eqref{uitau} into \eqref{convolutionct} and exploiting the fact that $\bm{\zeta}_i(t)= (\tilde{\bm{\Gamma}}^{-1}\otimes \mathbf{I}_{n_u}) \mathbf{U}_i(t)$, we find
\begin{align}
    \mathbf{x}_i(t) &= \int_0^\infty \mathbf{H}(\tau)\mathbf{u}_i(t-\tau)\textnormal{d}\tau \notag \\
    \label{secondline}
    &= \int_0^\infty \mathbf{H}(\tau) (\bm{\eta}^\top(\tau)\otimes \mathbf{I}_{n_u})\textnormal{d}\tau \mathbf{U}_i(t) \\
    &= \sum_{r=0}^{2M} \mathbf{H}_{\textnormal{MS}}(rh)\mathbf{u}_i(t-rh), \notag
\end{align}
where the second equality is due to the mixed-product property of the Kronecker product $(\bm{\varphi}^\top(\tau)\otimes \mathbf{I}_{n_u})(\tilde{\bm{\Gamma}}^{-1}\otimes \mathbf{I}_{n_u})=(\bm{\eta}^\top(\tau)\otimes \mathbf{I}_{n_u})$, with $\bm{\eta}(\tau)$ being defined in Definition \ref{definitionmultisine}. The last equality results from the blockwise multiplication of the matrix integral in \eqref{secondline} with $\mathbf{U}_i(t)$.

\textit{Proof of (2).} Let the matrix formed by the coefficients $\mathbf{H}_{\textnormal{MS}}(kh)$ be defined as 
\begin{equation}
\bm{\mathcal{H}}_{\textnormal{MS}}:=\big[\mathbf{H}_{\textnormal{MS}}(0),\mathbf{H}_{\textnormal{MS}}(h),\dots,\mathbf{H}_{\textnormal{MS}}(2Mh)\big]^\top,    
\end{equation}
and let the matrix of FRF evaluations be defined as
\begin{align}
    \bm{\mathcal{G}}_{\textnormal{MS}}&: = \big[\mathbf{G}_{\textnormal{MS}}(1), \mathbf{G}_{\textnormal{MS}}(e^{-\mathrm{i}\omega_1 h}), \mathbf{G}_{\textnormal{MS}}(e^{\mathrm{i}\omega_1 h}), \notag \\
\label{hatgf}
    &\dots, \mathbf{G}_{\textnormal{MS}}(e^{-\mathrm{i}\omega_M h}), \mathbf{G}_{\textnormal{MS}}(e^{\mathrm{i}\omega_M h})\big]^\top.
\end{align}
Based on these definitions and \eqref{definitionGMS}, we find that $\bm{\mathcal{G}}_{\textnormal{MS}}^\top = \bm{\mathcal{H}}_{\textnormal{MS}}^\top (\tilde{\bm{\Gamma}}\otimes \mathbf{I}_{n_u})$. Now, $\bm{\mathcal{H}}_{\textnormal{MS}}^\top = \int_0^\infty \mathbf{H}(\tau) (\bm{\eta}^\top(\tau)\otimes \mathbf{I}_{n_u})\textnormal{d}\tau$, which implies $\bm{\mathcal{G}}_{\textnormal{MS}}^{\top} = \int_0^\infty \mathbf{H}(\tau) (\bm{\varphi}^\top(\tau)\otimes \mathbf{I}_{n_u})\textnormal{d}\tau$. By inspection of each $n_y\times n_u$ block matrix forming $\bm{\mathcal{G}}_{\textnormal{MS}}$, we obtain $\mathbf{G}_{\textnormal{MS}}(1)= \int_0^\infty \mathbf{H}(\tau)\textnormal{d}\tau = \mathbf{G}(0)$, and $\mathbf{G}_{\textnormal{MS}}(e^{\pm\mathrm{i}\omega_\ell h})= \int_0^\infty \mathbf{H}(\tau)e^{\mp\mathrm{i}\omega_\ell\tau}\textnormal{d}\tau = \mathbf{G}(\pm\mathrm{i}\omega_\ell)$. \hfill $\square$

\begin{remark}
    The multisine equivalent discrete-time model in Definition \ref{definitionmultisine} is not unique and needs not to be defined as a causal model. In fact, a noncausal FIR model may also describe a resulting frequency response estimate that is equal to the continuous-time system at the input frequencies $\omega=0$ and $\omega=\pm\omega_\ell$ for $\ell=1,\dots,M$. To analyze such case, we consider the following FIR model:
    \begin{equation}
    \mathbf{G}_{\textnormal{MS},n_c}(z)= \sum_{k=n_c}^{2M+n_c} \mathbf{H}_{\textnormal{MS},n_c}(kh) z^{-k}, \notag 
    \end{equation}
    where $n_c\in\mathbb{Z}$, and where $\mathbf{H}_{\textnormal{MS},n_c}(kh)$ is defined similarly to~\eqref{hms} but for $k=n_c,\dots,2M+n_c$, and where $\bm{\eta}_k(\tau)$ is the $(k+1-n_c)$-th entry of the vector solving $\bm{\eta}^\top(\tau)\tilde{\bm{\Gamma}}=\bm{\varphi}^\top(\tau)\mathbf{D}_{n_c}$, where $\bm{\varphi}(\tau)$ and $\tilde{\bm{\Gamma}}$ are given by \eqref{varphi} and \eqref{gammatilde} respectively, and $\mathbf{D}_{{n_c}} = \textnormal{diag}(1,e^{\mathrm{i}n_c\omega_1 h},e^{-\mathrm{i}n_c\omega_1 h},\dots,e^{\mathrm{i}n_c\omega_M h},e^{-\mathrm{i}n_c\omega_M h})$. Via the same steps as in Lemma \ref{lem1}, it can be shown that $\mathbf{x}_i(kh) = \sum_{r=n_c}^{2M+n_c} \mathbf{H}_{\textnormal{MS},n_c}(rh)\mathbf{u}_i(t-rh)$, $\mathbf{G}(0)=\mathbf{G}_{\textnormal{MS},n_c}(1)$, and $\mathbf{G}(\pm \mathrm{i}\omega_\ell)=\mathbf{G}_{\textnormal{MS},n_c}(e^{\pm \mathrm{i}\omega_\ell h})$ for $\ell=1,\dots,M$. Thus, in summary, different values of $n_c$ yield models with identical frequency response values at $\omega=0$, $\omega=\pm\omega_\ell$, $\ell=1,\dots,M$, but differ at other frequencies due to the distinct trigonometric interpolation imposed by $n_c$. For simplicity of our exposition, we will consider $n_c=0$ for the remainder of this paper.
\end{remark}

An important observation from Lemma \ref{lem1} is that the multisine equivalent discrete-time formulation allows for an exact representation of any continuous-time system $\mathbf{G}_0(p)$ as a discrete-time FIR model \eqref{definitionGMS} for such type of inputs. This result holds even when the Nyquist-Shannon criterion is not satisfied. The exact discrete-time representation in Lemma \ref{lem1} leads to a straightforward least-squares estimation procedure in Section \ref{sub:statisticalpropertiesnonparametric} to estimate the multisine equivalent model.

\subsection{Least-squares estimator of matrix coefficients and its properties}
\label{sub:statisticalpropertiesnonparametric}

Since the system $\mathbf{G}_0(p)$ is unknown, the matrix coefficients $\mathbf{H}_{\textnormal{MS}}(kh), k=0,1,\dots,2M$ must be estimated from data. The following least-squares problem is proposed to estimate these coefficients, collected in $\bm{\mathcal{H}}_{\textnormal{MS}} \in \mathbb{R}^{(2M+1)n_u \times n_y}$:
\begin{equation}
\label{HMS}
\hat{\bm{\mathcal{H}}}_{\textnormal{MS}}\hspace{-0.06cm} =\hspace{-0.05cm}\underset{\bm{\mathcal{H}}_{\textnormal{MS}}}{\arg\min}\hspace{-0.015cm}\sum_{i=1}^m \hspace{-0.015cm}\sum_{k=1}^N\big\|\mathbf{y}_i\hspace{-0.02cm}(\hspace{-0.01cm}kh\hspace{-0.01cm})\hspace{-0.01cm}-\hspace{-0.01cm}\bm{\mathcal{H}}_{\textnormal{MS}}^\top\hspace{-0.03cm}\mathbf{U}_{\hspace{-0.02cm}i}\hspace{-0.03cm}(\hspace{-0.01cm}kh\hspace{-0.01cm})\big\|^2_{\bm{\Sigma}^{-1}}\hspace{-0.03cm}.
\end{equation}
Computing the matrix gradient of the cost function above with respect to $\bm{\mathcal{H}}_{\textnormal{MS}}$ and solving for $\bm{\mathcal{H}}_{\textnormal{MS}}$ gives the least-squares estimator
\begin{equation}
\label{ls}
    \hat{\bm{\mathcal{H}}}_{\textnormal{MS}} \hspace{-0.07cm}=\hspace{-0.09cm} \left[\sum_{k=1}^N \bm{\mathcal{U}}\hspace{-0.02cm}(kh) \bm{\mathcal{U}}^{\hspace{-0.02cm}\top}\hspace{-0.05cm}(kh) \hspace{-0.02cm}\right]^{\hspace{-0.06cm}-\hspace{-0.02cm}1}\hspace{-0.1cm}\sum_{k=1}^N  \bm{\mathcal{U}}\hspace{-0.02cm}(kh)\bm{\mathcal{Y}}^{\hspace{-0.02cm}\top}\hspace{-0.05cm}(kh),
\end{equation}
where we have defined the input and output matrices
\begin{align}
\label{inputmatrix}
    \bm{\mathcal{U}}(kh)&:= \Big[\mathbf{U}_1(kh), \dots, \mathbf{U}_m(kh)\Big], \\
    \bm{\mathcal{Y}}(kh)&:= \Big[\mathbf{y}_1(kh), \dots, \mathbf{y}_m(kh)\Big].
\end{align}
The following nonparametric estimator for the continuous-time FRF follows from the multisine discrete-time equivalent model and its frequency-domain property detailed in Lemma \ref{lem1}:
\begin{equation}
\label{frf}
    \hat{\mathbf{G}}(\mathrm{i}\omega) = \hat{\bm{\mathcal{H}}}_{\textnormal{MS}}^\top\big(\bm{\Gamma}(e^{\mathrm{i}\omega h})\otimes \mathbf{I}_{n_u}\big),
\end{equation}
where $\bm{\Gamma}(e^{\mathrm{i}\omega h})= [1, e^{-\mathrm{i}\omega h}, \dots, e^{-2M \mathrm{i} \omega h}]^\top$. Note that, by construction, the proposed estimator satisfies the conjugacy property $\overline{\hat{\mathbf{G}}(\mathrm{i}\omega)} = \hat{\mathbf{G}}(-\mathrm{i}\omega)$.

The explicit experimental conditions required for the well-posedness of this estimator and its unbiasedness are presented in Theorem \ref{thm2}. 
\begin{theorem}
	\label{thm2}
	Assume that $m\geq n_u$ experiments performed on the system described in \eqref{system} are conducted using multisines of the form \eqref{input}, designed under Assumption \ref{assumption1}. Furthermore, assume that Assumption \ref{assumption2} holds, the output of each experiment is in a stationary regime, and the number of data samples $N$ is greater than $2M$. Then, the least-squares estimator in \eqref{ls} is well-defined, and its associated frequency response estimator in Equation \eqref{frf} is unbiased at the frequencies $\omega = 0, \pm\omega_1,\pm\omega_2,\dots,\pm \omega_M$.
\end{theorem}

\textit{Proof.} We first prove that the least-squares estimator in \eqref{ls} is well-defined for $N\geq 2M+1$. This will occur if and only if its associated normal matrix is nonsingular. The matrix $\bm{\mathcal{U}}(kh)$ in \eqref{inputmatrix} can be written as
\begin{equation}
\label{Ukh}
    \bm{\mathcal{U}}(kh)=(\tilde{\bm{\Gamma}}\otimes \mathbf{I}_{n_u}) \bm{\mathcal{Z}}(kh),
\end{equation}
where the matrix
\begin{equation}
\label{zetamatrix}
\bm{\mathcal{Z}}(kh)=[\bm{\zeta}_1(kh),\dots, \bm{\zeta}_m(kh)]\in \mathbb{C}^{n_u(2M+1)\times m}    
\end{equation}
is formed by evaluating \eqref{zeta} at $t=kh$ for $i=1,\dots,m$, and $\tilde{\bm{\Gamma}}$ is the nonsingular matrix given in \eqref{gammatilde}. Thus, the normal matrix in \eqref{ls} is nonsingular if and only if 
\begin{equation}
\label{z}
    \mathbf{Z}:= \sum_{k=1}^N \hspace{-0.03cm}\bm{\mathcal{Z}}(kh) \bm{\mathcal{Z}}^\hop(kh)
\end{equation}
is nonsingular. Let $\mathbf{w}\in \mathbb{C}^{n_u(2M+1)}$ be arbitrary. We have $\mathbf{w}^\hop \mathbf{Z} \mathbf{w}=0$ if and only if $\bm{\mathcal{Z}}^\hop(kh)\mathbf{w}=0$ for all $k=1,\dots,N$. In particular, $\mathbf{Kw}=\mathbf{0}$, where $\mathbf{K}=[\bm{\mathcal{Z}}(h),\dots,\bm{\mathcal{Z}}([2M+1]h)]^\hop$. The rectangular matrix $\mathbf{K}$ can be decomposed as $\mathbf{K} = (\tilde{\bm{\Gamma}}\otimes \mathbf{I}_m) \mathbf{D}$, where $\mathbf{D}=\textnormal{blkdiag}\big( \hspace{-0.02cm}\mathbf{A}_0,\hspace{-0.03cm} e^{- \hspace{-0.02cm}\mathrm{i}\omega_1 \hspace{-0.02cm} h} \hspace{-0.02cm}\overline{\mathbf{A}_1},e^{ \hspace{-0.03cm}\mathrm{i}\omega_1 \hspace{-0.02cm} h} \hspace{-0.03cm}\mathbf{A}_1, \dots,$ $e^{\hspace{-0.03cm}-\mathrm{i}\omega_{M} \hspace{-0.02cm}h} \hspace{-0.02cm}\overline{\mathbf{A}_M},$ $e^{\mathrm{i}\omega_M \hspace{-0.02cm} h} \hspace{-0.03cm}\mathbf{A}_{ \hspace{-0.02cm}M}\big)$, with the $\mathbf{A}_\ell$ matrices being defined in \eqref{amatrices}. Since $\mathbf{A}_\ell$ $(\ell=0,1,\dots, m)$ have full column rank due to Assumption \ref{assumption1}, it follows that $\mathbf{D}$, and consequently $\mathbf{K}$, also have full column rank. Hence, if $\mathbf{Kw}=\mathbf{0}$, then $\mathbf{w}=\mathbf{0}$ and we conclude that $\mathbf{Z}$ is positive definite. This implies that the least-squares estimator \eqref{ls} is well-defined for $N\geq 2M+1$.

To establish unbiasedness, we compute the expected value of the least-squares estimator in \eqref{ls} as
\begin{align}
    \mathbb{E}\left\{\hat{\bm{\mathcal{H}}}_{\textnormal{MS}}\right\} =& (\tilde{\bm{\Gamma}}^{-\hop} \otimes \mathbf{I}_{n_u})\left[\sum_{k=1}^N \bm{\mathcal{Z}}(kh) \bm{\mathcal{Z}}^\hop(kh) \right]^{-1} \notag \\
    \label{ls2}
    &\hspace{-1cm}\times \sum_{k=1}^N \bm{\mathcal{Z}}(kh) \big[\mathbf{x}_1(kh), \dots, \mathbf{x}_m(kh)\big]^\top,
\end{align}
where we have exploited the representation of $\bm{\mathcal{U}}(kh)$ in \eqref{Ukh} and the fact that $\mathbf{v}_i(kh)$ in \eqref{output} has zero mean. Furthermore, due to the stationarity assumption on the output of each experiment we can write
\begin{equation}
\label{xkh}
    \mathbf{x}_i(kh)=\bm{\mathcal{G}}_0^\hop\bm{\zeta}_i(kh),
\end{equation}
where
\begin{align}
    \bm{\mathcal{G}}_0: = \big[\mathbf{G}_0(0)&, \mathbf{G}_0(-\mathrm{i}\omega_1), \mathbf{G}_0(\mathrm{i}\omega_1), \notag \\
    &\dots, \mathbf{G}_0(-\mathrm{i}\omega_M), \mathbf{G}_0(\mathrm{i}\omega_M)\big]^\top. \notag
\end{align}
Therefore, inserting \eqref{xkh} into \eqref{ls2}, we conclude that $(\tilde{\bm{\Gamma}}^\hop \otimes \mathbf{I}_{n_u})\mathbb{E}\{\hat{\bm{\mathcal{H}}}_{\textnormal{MS}}\}=\bm{\mathcal{G}}_0$, i.e., $\mathbb{E}\{\hat{\bm{\mathcal{H}}}^\top_{\textnormal{MS}}\}(\tilde{\bm{\Gamma}} \otimes \mathbf{I}_{n_u})=\bm{\mathcal{G}}_0^\hop$. By representing $\tilde{\bm{\Gamma}}$ in \eqref{gammatilde} in terms of $\bm{\Gamma}(e^{\mathrm{i}\omega h})$, we find the identities $\mathbb{E}\{\hat{\mathbf{G}}(0)\}=\mathbb{E}\{\hat{\bm{\mathcal{H}}}^\top_{\textnormal{MS}}\}(\bm{\Gamma}(e^{\mathrm{i}0 h}) \otimes \mathbf{I}_{n_u})=\mathbf{G}_0(0)$, and $\mathbb{E}\{\hat{\mathbf{G}}(\pm \mathrm{i}\omega_\ell)\}=\mathbb{E}\{\hat{\bm{\mathcal{H}}}^\top_{\textnormal{MS}}\}(\bm{\Gamma}(e^{\pm\mathrm{i}\omega_\ell h}) \otimes \mathbf{I}_{n_u})=\mathbf{G}_0(\pm \mathrm{i}\omega_\ell )$ for $\ell=1,\dots,M$. \hfill $\square$

The least-squares estimator of the multisine equivalent discrete-time model leads to unbiased estimates of the frequency response of the continuous-time system at the frequencies of interest. This result is valid for $N\geq 2M+1$ samples if the transients are removed from the data, or is valid asymptotically in the sample size if they are not. For the frequencies that are not excited by the multisine input, the least-squares method in \eqref{frf} delivers a trigonometric interpolation of the frequency response that is generally reliable if the frequencies that are excited are sufficiently densely distributed around the bandwidth of interest. However, caution must be exercised for the slow-sampling scenario, as the frequencies above the Nyquist frequency $\pi/h$ are aliased towards the fundamental frequency band $[-\pi/h,\pi/h)$ due to the periodicity of $\bm{\Gamma}(e^{\mathrm{i}\omega h})$ in \eqref{frf}. As a result, the estimated frequency response near these aliased frequencies may not accurately reflect the true continuous-time FRF within the fundamental frequency band.

The second result concerns the precision of the least-squares estimator. For a particular case where it is shown that the frequency response estimates are uncorrelated, we will require for the input to not exhibit spectral leakage, which means that only integer periods of the input signal must be used for identification.
\begin{assumption}[No spectral leakage]
\label{assumption3}
    The sampling period and sample size are such that $Nh$ is a multiple of the least common multiple of $\{2\pi/\omega_\ell\}_{\ell=1}^M$.
\end{assumption}

In the same way as in \eqref{hatgf}, we denote the matrix that stacks the frequency response function estimates as $\hat{\bm{\mathcal{G}}}_{\textnormal{MS}}$. The covariance of the least-squares estimator of the frequency response function can be computed in closed-form for finite samples, as seen next.

\begin{theorem}
	\label{thm3}
	Consider the least-squares estimator in \eqref{ls}, and its associated frequency response function estimator in~\eqref{frf}. Assume that Assumptions \ref{assumption1} and \ref{assumption2} hold, and that the system output is in a stationary regime. Then, the following result holds for any sample size $N> 2M$:
	\begin{equation}
    \label{covariancegf}
	\textnormal{Cov}\left\{\textnormal{vec}\{\hat{\bm{\mathcal{G}}}_{\textnormal{MS}}\}\right\} =  \bm{\Sigma}\otimes \mathbf{Z}^{-1},
	\end{equation}
	where $\mathbf{Z}$ is given by \eqref{z}. Furthermore, if Assumption \ref{assumption3} also holds, then the frequency response estimates $\hat{\mathbf{G}}(0), \hat{\mathbf{G}}(-\mathrm{i}\omega_1), \hat{\mathbf{G}}(\mathrm{i}\omega_1),\dots,  \hat{\mathbf{G}}(-\mathrm{i}\omega_M), \hat{\mathbf{G}}(\mathrm{i}\omega_M)$ are mutually uncorrelated.
\end{theorem}

\textit{Proof.}  We begin by computing $\hat{\bm{\mathcal{G}}}_{\textnormal{MS}}-\mathbb{E}\{\hat{\bm{\mathcal{G}}}_{\textnormal{MS}}\}$ in terms of the noise components $\mathbf{v}_i(kh)$. From \eqref{ls2}, we find
    \begin{align}
        \hat{\bm{\mathcal{G}}}_{\textnormal{MS}}-\mathbb{E}\{\hat{\bm{\mathcal{G}}}_{\textnormal{MS}}\} &= (\tilde{\bm{\Gamma}}^{\hop} \otimes \mathbf{I}_{n_u})\big(\hat{\bm{\mathcal{H}}}_{\textnormal{MS}}-\mathbb{E}\{\hat{\bm{\mathcal{H}}}_{\textnormal{MS}}\} \big) \notag \\
        \label{intermediate}
        &= \mathbf{Z}^{-1} \sum_{k=1}^N \bm{\mathcal{Z}}(kh)\bm{\mathcal{V}}^\top(kh),
    \end{align}
where $\mathbf{Z}$ is defined in \eqref{z}, and the noise component $\bm{\mathcal{V}}(kh)$ is given by $\bm{\mathcal{V}}(kh)=[\mathbf{v}_1(kh),\dots,\mathbf{v}_m(kh)]$. In vectorized form, using the identity $\textnormal{vec}\{\mathbf{A}\mathbf{B}\}=(\mathbf{I}_m\otimes \mathbf{A})\textnormal{vec}\{\mathbf{B}\}$ where $\mathbf{B}$ has $m$ columns, \eqref{intermediate} is equivalent to
\begin{align}
    \textnormal{vec}&\{\hat{\bm{\mathcal{G}}}_{\textnormal{MS}}\}-\textnormal{vec}\{\mathbb{E}\{\hat{\bm{\mathcal{G}}}_{\textnormal{MS}}\}\} =\notag \\
\label{combined}
    &(\mathbf{I}_{n_y}\otimes \mathbf{Z})^{-1}\sum_{k=1}^N (\mathbf{I}_{n_y}\otimes \bm{\mathcal{Z}}(kh))\textnormal{vec}\{\bm{\mathcal{V}}^\top(kh)\}.
\end{align}
On the other hand, the covariance of the noise vector $\textnormal{vec}\{\bm{\mathcal{V}}^\top(kh)\}$ is
\begin{equation}
\textnormal{Cov}\hspace{-0.05cm}\left\{\hspace{-0.04cm}\textnormal{vec}\{\bm{\mathcal{V}}^\top\hspace{-0.04cm}(kh)\}[\textnormal{vec}\{\bm{\mathcal{V}}^\top\hspace{-0.05cm}(lh)\}]^\top\right\} \hspace{-0.06cm}=\hspace{-0.05cm}(\bm{\Sigma}\hspace{0.02cm}\otimes\hspace{0.02cm} \mathbf{I}_m)\delta_K(k-l) , \notag 
\end{equation}
which, when combined with \eqref{combined}, implies that
\begin{align}
    \textnormal{Cov}\big\{\textnormal{vec}\{\hat{\bm{\mathcal{G}}}_{\textnormal{MS}}\}\big\} &= (\mathbf{I}_{n_y}\otimes \mathbf{Z})^{-1} \sum_{k=1}^N \big[(\mathbf{I}_{n_y}\otimes \bm{\mathcal{Z}}(kh)) \notag \\
    \times\hspace{0.02cm}(\bm{\Sigma}\hspace{0.02cm}\otimes\hspace{0.02cm}& \mathbf{I}_{m})(\mathbf{I}_{n_y}\otimes \bm{\mathcal{Z}}^\hop(kh))\big] (\mathbf{I}_{n_y}\otimes \mathbf{Z})^{-1} \notag \\
    &=(\mathbf{I}_{n_y}\otimes \mathbf{Z})^{-1}(\bm{\Sigma}\otimes \mathbf{Z}) (\mathbf{I}_{n_y}\otimes \mathbf{Z})^{-1} \notag \\
    &=\bm{\Sigma}\otimes \mathbf{Z}^{-1}, \notag 
\end{align}
where we have used the bilinearity property of the Kronecker product, the mixed-product property $(\mathbf{A}\otimes \mathbf{B})(\mathbf{C}\otimes \mathbf{D})=(\mathbf{AC}\otimes \mathbf{BD})$, and the matrix inverse property $(\mathbf{A}\otimes \mathbf{B})^{-1}=\mathbf{A}^{-1}\otimes \mathbf{B}^{-1}$.
 
What is left to prove is the uncorrelatedness of the frequency response estimates at distinct frequencies. The frequency response estimates $\hat{\mathbf{G}}(0)$, $\hat{\mathbf{G}}(-\mathrm{i}\omega_1)$, $\hat{\mathbf{G}}(\mathrm{i}\omega_1),\dots$, $ \hat{\mathbf{G}}(-\mathrm{i}\omega_M)$, $\hat{\mathbf{G}}(\mathrm{i}\omega_M)$ are mutually uncorrelated if and only if the matrix $\mathbf{Z}$ is block diagonal, with each diagonal matrix being of size $n_u\times n_u$. If we denote $\mathbf{E}_c\in\mathbb{R}^{n_u(2M+1)\times n_u}, c\in \{1,2,\dots,2M+1\}$ as the matrix formed by columns $n_u(c-1)+1$ to $n_u c$ of the identity matrix $\mathbf{I}_{n_u(2M+1)}$, then
\begin{align}
    \mathbf{E}_r^\top \mathbf{Z}\mathbf{E}_c &= \sum_{i=1}^m \sum_{k=1}^N \mathbf{E}_r^\top \bm{\zeta}_i(kh)\bm{\zeta}_i^\hop(kh) \mathbf{E}_c \notag \\
    &=C\sum_{i=1}^m \Big(\mathbf{a}_{\lfloor \frac{r}{2} \rfloor,i}\odot e^{\mathrm{i}(-1)^{r}\bm{\phi}_{\lfloor \frac{r}{2} \rfloor,i}}  \Big) \notag \\
    &\times \Big(\mathbf{a}^\top_{\lfloor \frac{c}{2} \rfloor,i}\odot e^{\mathrm{i}(-1)^{c+1}\bm{\phi}^\top_{\lfloor \frac{c}{2} \rfloor,i}} \Big)\sum_{k=1}^N e^{\mathrm{i}kh \bar{\omega}} \notag \\
    \label{powerunity}
    &=  \mathcal{M}_{rc}(\mathbf{A}_{\lfloor \frac{r}{2} \rfloor},\mathbf{A}_{\lfloor \frac{c}{2} \rfloor}) \sum_{k=1}^N e^{\mathrm{i}kh \bar{\omega}},
\end{align}
where $C$ is a scalar constant, $\bar{\omega}=\omega_{\lfloor \frac{r}{2} \rfloor}(-1)^r -\omega_{\lfloor \frac{c}{2} \rfloor}(-1)^c$, and where we have denoted $\omega_0=0$ and $\bm{\phi}_0=\mathbf{0}$. The matrix $\mathcal{M}_{rc}(\mathbf{A}_{\lfloor \frac{r}{2} \rfloor},\mathbf{A}_{\lfloor \frac{c}{2} \rfloor})$ is given by
\begin{equation}
    \mathcal{M}_{rc}(\mathbf{A}_{\lfloor \frac{r}{2} \rfloor}\hspace{-0.02cm},\hspace{-0.03cm}\mathbf{A}_{\lfloor \frac{c}{2} \rfloor}\hspace{-0.02cm})\hspace{-0.03cm}=\hspace{-0.06cm}\begin{cases}
        \mathbf{A}_{\lfloor \frac{r}{2} \rfloor}^\top \overline{\mathbf{A}_{\lfloor \frac{c}{2} \rfloor}}, & \hspace{-0.05cm}\textnormal{if }r,c \textnormal{ are even},\\
        \mathbf{A}_{\lfloor \frac{r}{2} \rfloor}^\hop \mathbf{A}_{\lfloor \frac{c}{2} \rfloor}, & \hspace{-0.05cm}\textnormal{if }r,c \textnormal{ are odd},\\
        \mathbf{A}_{\lfloor \frac{r}{2} \rfloor}^\top \mathbf{A}_{\lfloor \frac{c}{2} \rfloor}, & \hspace{-0.05cm}\textnormal{if }r \textnormal{ is even},c \textnormal{ is odd},\\
        \mathbf{A}_{\lfloor \frac{r}{2} \rfloor}^\hop \overline{\mathbf{A}_{\lfloor \frac{c}{2} \rfloor}}, & \hspace{-0.05cm}\textnormal{if }r \textnormal{ is odd},c \textnormal{ is even}.
    \end{cases} \notag 
\end{equation}
Since $Nh$ is assumed to be a multiple of the least common multiple of $\{2\pi/\omega_\ell\}_{\ell=1}^M$ due to Assumption \ref{assumption3}, for each $\ell\in \{1,\dots,M\}$ there exists an integer $d_\ell$ such that $Nh=2\pi d_\ell/\omega_\ell$. Thus, $\bar{\omega}$ can be written as $\bar{\omega}=2\pi n_{rc}/(Nh)$, where $n_{rc}=d_{\lfloor \frac{r}{2} \rfloor}(-1)^r-d_{\lfloor \frac{c}{2} \rfloor}(-1)^c$ is a non-zero integer that is not a multiple of $N$ due to Assumption \ref{assumption2}.

After substituting $\bar{\omega}=2\pi n_{rc}/(Nh)$ in \eqref{powerunity}, the resulting complex-exponential sum in \eqref{powerunity} becomes the sum of the $n_{rc}$-th power of the $N$-th roots of unity, which is zero whenever $n_{rc}$ is not a multiple of $N$ (see, e.g., \cite[Ex. 3, Ch. 3]{ledermann1962complex}). Consequently, $\mathbf{E}_r^\top \mathbf{Z}\mathbf{E}_c=\mathbf{0}$ for $r\neq c$, which proves that $\mathbf{Z}$ is block diagonal with the appropriate block dimensions. \hfill $\square$

\begin{remark}
    If the spectral leakage condition stated in Assumption \ref{assumption3} is not met, it is still possible to show that the frequency response estimates are asymptotically uncorrelated as the sample size tends to infinity. This observation can be derived with the same approach as in \eqref{powerunity}, but its formal proof is beyond the scope of this paper.
\end{remark}

Since Theorems \ref{thm2} and \ref{thm3} are valid for any sample size greater than $2M$, they enable the exploration of limit results as $m$ or $N$ approach infinity. It is possible to prove that the least-squares estimator for $\bm{\mathcal{G}}_0$ is weakly consistent (i.e., it converges in probability to $\bm{\mathcal{G}}_0$) under mild input conditions as the number of experiments $m$, or sample size $N$, approaches infinity. Corollary \ref{coroconsistency} formalizes this result, extending the statistical analysis of discrete-time parametric identification with multiple data sets \cite{markovsky2015identification,vasquez2022consistency} to the continuous-time MIMO case with multisine inputs in a nonparametric framework.

\begin{corollary}
\label{coroconsistency}
Under the experimental conditions in Theorem \ref{thm2}, the least-squares estimator of the frequency response function, $\hat{\bm{\mathcal{G}}}_{\textnormal{MS}}$, is a weakly consistent estimator of $\bm{\mathcal{G}}_0$ as $N \to \infty$ with a fixed number of experiments $m$.

Moreover, under the same experimental conditions and assuming Assumption \ref{assumption3}, $\hat{\bm{\mathcal{G}}}_{\textnormal{MS}}$ is also weakly consistent for $\bm{\mathcal{G}}_0$ as $m\to\infty$ with a fixed sample size $N$ per experiment, provided that the following conditions simultaneously hold:
\begin{enumerate}[label=\arabic*.]
    \item There exists $\gamma>0$ such that, for all $\ell = 0, 1, \dots, M$ and $i\neq j$ with $1\leq i\neq j\leq n_u$, the columns $\mathbf{c}_\ell^{(i)}$ of $\mathbf{A}_\ell$ ($\ell = 0, 1, \dots, M$) satisfy the mutual coherence bound
\begin{equation}
\label{mutualcoherence}
    (n_u-1)\lim_{m\to\infty} \frac{|\langle \mathbf{c}_\ell^{(i)}, \mathbf{c}_\ell^{(j)} \rangle|}{\|\mathbf{c}_\ell^{(i)}\|\|\mathbf{c}_\ell^{(j)}\|}<1-\gamma.
\end{equation}
\item For all $i=1,\dots,n_u$ and $\ell=0,1,\dots,M$, $\|\mathbf{c}_\ell^{(i)}\|\to \infty$ as $m\to\infty$.
\end{enumerate}
\end{corollary}

\textit{Proof.} Since the least-squares estimator is unbiased due to Theorem \ref{thm2}, Theorem 8.2 (page 54) of \cite{lehmann1998theory} implies that, to prove each weak consistency property, it is sufficient to show that the variance of each element of $\hat{\bm{\mathcal{G}}}_{\textnormal{MS}}$ converges to zero as $N\to\infty$ or as $m\to\infty$. From \eqref{covariancegf}, this is guaranteed if $\mathbf{Z}^{-1}\to\mathbf{0}$, or equivalently, if $\lambda_{\textnormal{min}}(\mathbf{Z})\to\infty$. By the block matrix version of Ger\v{s}gorin's circle theorem \cite[Chap. 6.1.P17]{Horn2012}, each eigenvalue of $\mathbf{Z}$ lies within the set
\begin{equation}
\label{theset}
    \bigcup_{r=1}^{2M+1}\bigcup_{j=1}^{n_u} \hspace{-0.04cm}\bigg\{z\in\mathbb{R}\colon z\hspace{-0.02cm}\geq\hspace{-0.02cm} \lambda_j^{(r)}\hspace{-0.07cm}-\hspace{-0.3cm}\sum_{1\leq c\neq r \leq 2M+1} \hspace{-0.3cm} \|\mathbf{E}_r^\top \mathbf{Z}\mathbf{E}_c\| \hspace{-0.03cm}\bigg\},
\end{equation}
where $\lambda_j^{(r)}$ denotes the $j$th eigenvalue of $\mathbf{E}_r^\top \mathbf{Z}\mathbf{E}_r$. It follows from \eqref{powerunity} that 
\begin{align}
    \lambda_j^{(r)}&\geq  N \min_{0\leq \ell \leq 2M}\sigma_{\textnormal{min}}^2(\mathbf{A}_\ell),  \notag \\
    \sum_{1\leq c\neq r \leq 2M+1} \hspace{-0.3cm}\|\mathbf{E}_r^\top \mathbf{Z}\mathbf{E}_c\| &\leq \hspace{-0.2cm}\sum_{1\leq c\neq r \leq 2M+1} \hspace{-0.35cm} \frac{\|\mathcal{M}_{rc}(\mathbf{A}_{\lfloor \frac{r}{2} \rfloor}\hspace{-0.03cm},\hspace{-0.015cm}\mathbf{A}_{\lfloor \frac{c}{2} \rfloor})\| }{|\sin(\frac{h\bar{\omega}}{2})|}. \notag 
\end{align}
Importantly, the lower bound on $\lambda_j^{(r)}$ is positive and increases with $N$, while the upper bound on the off-diagonal norm sum is finite and independent of $N$. Substituting these into \eqref{theset} yields a relaxed bound showing that all eigenvalues of $\mathbf{Z}$ grow at least linearly in $N$. Therefore, $\lambda_{\textnormal{min}}(\mathbf{Z}) \to \infty$ as $N \to \infty$.

For the weak consistency result when $m\to\infty$, we note that under Assumptions \ref{assumption1} to  \ref{assumption3}, the matrix $\mathbf{Z}$ is block-diagonal due to Theorem \ref{thm3}, with block-diagonal terms given by $N\mathbf{A}_0^\top \mathbf{A}_0$, $N\overline{\mathbf{A}_\ell^\hop \mathbf{A}_\ell}$, and $N\mathbf{A}_\ell^\hop \mathbf{A}_\ell$, for $\ell=1,\dots,M$. For simplicity we consider $N\mathbf{A}_0^\top \mathbf{A}_0$; the same reasoning applies to each of the other block-diagonal terms. After denoting the columns of $\mathbf{A}_0$ as $\mathbf{c}_0^{(i)}, i=1,\dots,n_u$, we consider the decomposition $\mathbf{A}_0 = \tilde{\mathbf{A}}_0 \mathbf{D}$, with $\mathbf{D}=\textnormal{diag}(\|\mathbf{c}_0^{(1)}\|,\dots,\|\mathbf{c}_0^{(n_u)}\|)$. By the Rayleigh quotient characterization of the smallest eigenvalue of $\mathbf{A}_0^\top \mathbf{A}_0$, we find
\begin{align}
    \lambda_{\textnormal{min}}(\mathbf{A}_0^{\top} \mathbf{A}_0)&=\lambda_{\textnormal{min}}(\mathbf{D}\tilde{\mathbf{A}}_0^{\top} \tilde{\mathbf{A}}_0\mathbf{D}) \notag \\
    &\geq \lambda_{\textnormal{min}}(\tilde{\mathbf{A}}_0^\top \tilde{\mathbf{A}}_0) \min_{1\leq i \leq n_u} \|\mathbf{c}_0^{(i)}\|^2 , \notag
\end{align}
where $\tilde{\mathbf{A}}_0^\top \tilde{\mathbf{A}}_0$ is a Gram matrix with unit diagonal entries. It follows from Ger\v{s}gorin's circle theorem that the smallest eigenvalue of $\tilde{\mathbf{A}}_0^\top \tilde{\mathbf{A}}_0$ is bounded by
\begin{equation}
    \lambda_{\textnormal{min}}(\tilde{\mathbf{A}}_0^\top \tilde{\mathbf{A}}_0) \geq 1 - (n_u-1)\max_{1\leq i\neq j \leq n_u} \frac{|\langle\mathbf{c}_0^{(i)},\mathbf{c}_0^{(j)}\rangle|}{\|\mathbf{c}_0^{(i)}\|\|\mathbf{c}_0^{(j)}\|}. \notag 
\end{equation}
The mutual coherence bound in Condition 1 then guarantees a uniform lower bound on $\lambda_{\textnormal{min}}(\tilde{\mathbf{A}}_0^\top \tilde{\mathbf{A}}_0)$ for large $m$, ensuring that the Gram matrix is nonsingular when $m\to\infty$. Moreover, since for large $m$ we have $\lambda_{\textnormal{min}}(\tilde{\mathbf{A}}_0^\top \tilde{\mathbf{A}}_0)\geq \gamma \min_{1\leq i \leq n_u} \|\mathbf{c}_0^{(i)}\|$, Condition 2 of the corollary statement yields $\lambda_{\textnormal{min}}(\tilde{\mathbf{A}}_0^\top \tilde{\mathbf{A}}_0)\to\infty$ as $m\to\infty$, which implies that $(N\mathbf{A}_0^\top \mathbf{A}_0)^{-1}\to \mathbf{0}$ as $m\to \infty$. Proceeding similarly with the remaining block-diagonal terms yields $\textnormal{Cov}\{\textnormal{vec}\{\hat{\bm{\mathcal{G}}}_{\textnormal{MS}}\}\} \to \mathbf{0}$ as $m \to \infty$, which completes the proof. \hfill $\square$

Corollary \ref{coroconsistency} shows that for weak consistency to hold as the number of identification experiments increases, it is sufficient to design the input frequency amplitudes so that they form a series that is not square summable. This condition allows the amplitudes to decay to zero when the number of experiments increases, provided the decay rate is no faster than $1/\sqrt{m}$. Condition~1 of Corollary \ref{coroconsistency} further requires that the columns of $\mathbf{A}_\ell$ be sufficiently linearly independent. This is quantified formally with the mutual coherence, commonly used in compressed sensing \cite{elad2010sparse} and sparse system identification \cite{parsa2023transformation}.

In summary, we have demonstrated  that the proposed least-squares estimator for the FRF at the input frequencies is unbiased, established an exact 
condition for the mutual uncorrelatedness of the frequency response estimates within the context of finite-duration experiments, and established explicit conditions for weak consistency as either the sample size or the number of experiments tends to infinity. A direct consequence of Theorem \ref{thm3} is that, when an integer period of the input signal is recorded and the additive measurement noise is Gaussian, the FRF estimates at each frequency are mutually statistically independent. This property enables both computationally and statistically efficient procedures for parametric system identification, as discussed in Section \ref{sec:parametric}. 

\subsection{Frequency domain interpretation}
\label{sub:frequency}
Due to the fact that multisine inputs have discrete frequency spectra, it is natural to consider a frequency-domain interpretation of the least-squares estimator in \eqref{ls}. Via Parseval's theorem \cite{oppenheim1997signals} and the identity $\hat{\bm{\mathcal{G}}}_{\textnormal{MS}} = (\tilde{\bm{\Gamma}}^{\hop}\otimes \mathbf{I}_{n_u})\hat{\bm{\mathcal{H}}}_{\textnormal{MS}}$, \eqref{ls} is equivalent to
\begin{equation}
\label{lsfreq}
    \hat{\bm{\mathcal{G}}}_{\textnormal{MS}} \hspace{-0.08cm}=\hspace{-0.1cm} \left[\sum_{n=1}^N \hspace{-0.05cm}\bm{\Psi}\hspace{-0.01cm}[e^{\mathrm{i}\frac{2\pi n}{N}}] \bm{\Psi}^{\hspace{-0.01cm}\hop}\hspace{-0.03cm}[e^{\mathrm{i}\frac{2\pi n}{N}}] \hspace{-0.02cm}\right]^{\hspace{-0.06cm}-\hspace{-0.02cm}1}\hspace{-0.1cm} \sum_{n=1}^N \hspace{-0.05cm} \bm{\Psi}\hspace{-0.01cm}[e^{\mathrm{i}\frac{2\pi n}{N}}]\bm{\Upsilon}^{\hspace{-0.01cm}\hop}\hspace{-0.01cm}[e^{\mathrm{i}\frac{2\pi n}{N}}]\hspace{0.02cm},
\end{equation}
where the discrete-time Fourier transforms (DTFTs) of $\bm{\mathcal{Z}}(kh)$ and $\bm{\mathcal{Y}}(kh)$ are respectively given by
\begin{equation}
    \bm{\Psi}\hspace{-0.02cm}[e^{\mathrm{i}\omega h}]\hspace{-0.05cm}= \hspace{-0.05cm}\sum_{k=1}^N \hspace{-0.03cm}\bm{\mathcal{Z}}(kh)e^{\hspace{-0.03cm}-\mathrm{i}k\omega h}, \quad \hspace{-0.12cm}\bm{\Upsilon}\hspace{-0.02cm}[e^{\mathrm{i}\omega h}]\hspace{-0.05cm}=\hspace{-0.05cm} \sum_{k=1}^N \hspace{-0.03cm}\bm{\mathcal{Y}}(kh)e^{\hspace{-0.02cm}-\mathrm{i}k\omega h}. \notag  
\end{equation}
The trigonometric interpolation imposed by the multisine discrete-time equivalent model also leads to a trigonometric interpolation of the FRF estimates, which in the frequency domain can be interpreted as the quotient of DTFTs of the input and output samples evaluated at each input frequency. This interpretation is provided in Theorem \ref{thmfrequency}.

\begin{theorem}
\label{thmfrequency}
    Consider the least-squares estimator in \eqref{ls}, and its associated frequency response function estimator in~\eqref{frf}. Assume that Assumptions \ref{assumption1} and \ref{assumption2} hold, and that the system output is in a stationary regime. Then, for $N> 2M$, the continuous-time frequency response estimator in \eqref{frf} is given by
    \begin{equation}
\label{frf2}
    \hat{\mathbf{G}}(\mathrm{i}\omega) = \hat{\bm{\mathcal{G}}}_{\textnormal{MS}}^\top\big(\overline{\tilde{\bm{\Gamma}}}^{-1}\bm{\Gamma}(e^{\mathrm{i}\omega h})\otimes \mathbf{I}_{n_u}\big),
\end{equation}
where the frequency response estimates $\hat{\bm{\mathcal{G}}}_{\textnormal{MS}}$ can be computed as in \eqref{lsfreq}. Furthermore, if Assumption \ref{assumption3} also holds, $\hat{\bm{\mathcal{G}}}_{\textnormal{MS}}$ is given by
\begin{equation}
\label{etfe}
 \hat{\bm{\mathcal{G}}}_{\textnormal{MS}} = \begin{bmatrix}
     \bm{\Upsilon}[e^{\mathrm{i}0}]\bm{\Xi}^\dagger [e^{\mathrm{i}0}]  \\
    \bm{\Upsilon}[e^{-\mathrm{i}\omega_1 h}]\bm{\Xi}^\dagger [e^{-\mathrm{i}\omega_1 h}] \\
    \bm{\Upsilon}[e^{\mathrm{i}\omega_1 h}] \bm{\Xi}^\dagger [e^{\mathrm{i}\omega_1 h}] \\
     \vdots \\
      \bm{\Upsilon}[e^{-\mathrm{i}\omega_M h}] \bm{\Xi}^\dagger [e^{-\mathrm{i}\omega_M h}] \\
    \bm{\Upsilon}[e^{\mathrm{i}\omega_M h}]\bm{\Xi}^\dagger [e^{\mathrm{i}\omega_M h}]
 \end{bmatrix}, 
\end{equation}
where $\bm{\Xi}[e^{\mathrm{i}\omega h}]$ and $\bm{\Upsilon}[e^{\mathrm{i}\omega h}]$ denote the DTFTs of $\bm{\mathcal{U}}(kh)$ and $\bm{\mathcal{Y}}(kh)$, respectively.
\end{theorem}

\textit{Proof}: The equivalent expression in \eqref{frf2} is direct by replacing $\hat{\bm{\mathcal{H}}}_{\textnormal{MS}} = (\tilde{\bm{\Gamma}}^{-\hop}\otimes \mathbf{I}_{n_u})\hat{\bm{\mathcal{G}}}_{\textnormal{MS}}$ into \eqref{frf}. For proving \eqref{etfe}, we require computing the discrete Fourier transform of $\bm{\mathcal{Z}}(kh)$. Due to Assumption \ref{assumption3}, $\bm{\Psi}[e^{\mathrm{i}\frac{2\pi n}{N}}]$ is an $N$-periodic sequence that can be computed from standard algebraic manipulations. One period of such transform is described by
\begin{equation}
\label{psitransform}
    \bm{\Psi}\hspace{-0.02cm}[e^{\mathrm{i}\frac{2\pi n}{N}}]= N \begin{bmatrix}
        \mathbf{A}_0^\top \delta_K(n \hspace{0.04cm}\textnormal{mod }N) \\
        \mathbf{A}_1^\top \delta_K((n-\frac{Nh\omega_1}{2\pi}) \hspace{0.04cm}\textnormal{mod }N) \\
        \mathbf{A}_1^\hop \delta_K((n+\frac{Nh\omega_1}{2\pi}) \hspace{0.04cm}\textnormal{mod }N) \\
        \vdots \\
        \mathbf{A}_M^\top \delta_K((n-\frac{Nh\omega_M}{2\pi}) \hspace{0.04cm}\textnormal{mod }N) \\ 
        \mathbf{A}_M^\hop \delta_K((n+\frac{Nh\omega_M}{2\pi}) \hspace{0.04cm}\textnormal{mod }N)
    \end{bmatrix},
\end{equation}
which leads to $\sum_{n=1}^N \hspace{-0.1cm}\bm{\Psi}[e^{\mathrm{i}\frac{2\pi n}{N}}\hspace{-0.02cm}] \bm{\Psi}^{\hop}[e^{\mathrm{i}\frac{2\pi n}{N}}\hspace{-0.02cm}]\hspace{-0.08cm}=\hspace{-0.08cm}N^2\textnormal{blkdiag}(\mathbf{A}_0^\top \hspace{-0.05cm}\mathbf{A}_0,$ $\overline{\mathbf{A}_1^\hop \mathbf{A}_1},\mathbf{A}_1^\hop \mathbf{A}_1,\dots,\overline{\mathbf{A}_M^\hop \mathbf{A}_M},\mathbf{A}_M^\hop \mathbf{A}_M)$, where the block diagonal terms are nonsingular due to Assumption \ref{assumption1}, and the terms outside the block diagonal are zero due to Assumption~\ref{assumption2}. Moreover,
\begin{equation}
    \sum_{n=1}^N \bm{\Psi}[e^{\mathrm{i}\frac{2\pi n}{N}}]\bm{\Upsilon}^{\hop}[e^{\mathrm{i}\frac{2\pi n}{N}}] = N \begin{bmatrix}
        \mathbf{A}_0^\top \bm{\Upsilon}^{\hop}[e^{\mathrm{i}0}] \\
        \mathbf{A}_1^\top \bm{\Upsilon}^{\hop}[e^{\mathrm{i}\omega_1 h}] \\
        \mathbf{A}_1^\hop \bm{\Upsilon}^{\hop}[e^{-\mathrm{i}\omega_1 h}] \\
        \vdots \\
        \mathbf{A}_M^\top \bm{\Upsilon}^{\hop}[e^{\mathrm{i}\omega_M h}] \\
        \mathbf{A}_M^\hop \bm{\Upsilon}^{\hop}[e^{-\mathrm{i}\omega_M h}]
    \end{bmatrix}. \notag
\end{equation}
In light of these expressions, we find that $\hat{\mathbf{G}}(0) = \bm{\Upsilon}[e^{\mathrm{i}0}]\mathbf{A}_0(N\mathbf{A}_0^\top\mathbf{A}_0)^{-1}$ and, for $\ell=1,\dots,M$, $\hat{\mathbf{G}}(\mathrm{i}\omega_\ell) = \bm{\Upsilon}[e^{\mathrm{i}\omega_\ell h}]\overline{\mathbf{A}_\ell}(N\mathbf{A}_\ell^\top\overline{\mathbf{A}_\ell})^{-1}$. On the other hand, from \eqref{psitransform} and the fact that $\bm{\mathcal{U}}(kh)=(\mathbf{1}_{2M+1}^\top \otimes \mathbf{I}_{n_u})\bm{\mathcal{Z}}(kh)$, it can be verified that the DTFT of the input matrix $\bm{\mathcal{U}}(kh)$, denoted by $\bm{\Xi}[e^{\mathrm{i}\omega h}]$, satisfies $\bm{\Xi}[e^{\mathrm{i}0}] = N\mathbf{A}_0^\top$, $\bm{\Xi}[e^{\mathrm{i}\omega_\ell h}] = N \mathbf{A}^\top_\ell$ and $\bm{\Xi}[e^{-\mathrm{i}\omega_\ell h}] = N \mathbf{A}^\hop_\ell$. Replacing these results into \eqref{lsfreq} leads to~\eqref{etfe}, concluding the proof. \hfill $\square$

Theorem \ref{thmfrequency} indicates that, when an integer period of the input signal is recorded, the least-squares estimator in \eqref{frf} provides an interpolation of the ETFE \cite{ljung1985estimation} for continuous-time systems using sampled data. In consequence, the statistical properties derived in Theorems \ref{thm2} and \ref{thm3} extend the results in \cite{ljung1985estimation} to a MIMO continuous-time system framework. Additionally, these theorems provide the exact computation of the covariance matrix of the ETFE, complementing prior asymptotic expressions for errors-in-variables cases derived in~\cite[Eq. 2-77]{pintelon2012system}.

\section{Direct and two-step parametric estimators}
\label{sec:parametric}
The finite-time properties stated in Theorems \ref{thm2} to \ref{thmfrequency} provide the essential tools towards the analysis of parametric estimators of the MIMO continuous-time system in \eqref{system} based on the multisine equivalent discrete-time FIR model in \eqref{frf}. The key idea behind the upcoming analysis is that the time-domain data from multiple experiments can be compressed without loss of statistical information. 

We consider a parametric model structure $\mathbf{G}(p,\bm{\theta})$ of $\mathbf{G}_0(p)$, where $\bm{\theta}\in \mathbb{R}^{n_{\theta}}$ contains the coefficients associated with, for example, the numerator and denominator polynomials of rational transfer functions in matrix fraction description form, the flexible and rigid-body modes in mechanical systems described in additive or modal form \cite{gonzalez2025statistically}, or continuous-time state-space descriptions. This parameter vector is assumed to be constrained to a known compact set $\mathcal{D}$. For the sequel, we define the matrix of frequency response evaluations of the parametric model as
\begin{align}
    \bm{\mathcal{G}}(\bm{\theta})&: = \big[\mathbf{G}(0,\bm{\theta}), \mathbf{G}(-\mathrm{i}\omega_1,\bm{\theta}), \mathbf{G}(\mathrm{i}\omega_1,\bm{\theta}), \notag \\
\label{Gftheta}
    &\hspace{0.5cm}\dots, \mathbf{G}(-\mathrm{i}\omega_M,\bm{\theta}), \mathbf{G}(\mathrm{i}\omega_M,\bm{\theta})\big]^\top.
    \end{align}
The main contributions in this section are the explicit relations between direct (i.e., one-step) and two-step parametric estimators $\hat{\bm{\theta}}$ describing the model $\mathbf{G}(p,\hat{\bm{\theta}})$ for the true continuous-time system $\mathbf{G}_0(p)$, and the implications of this result for computational methods and finite-sample statistical analyses. A key result on data compression is given next, that relates the least-squares estimator of the frequency response in \eqref{frf} with the sampled input and output data.

\begin{theorem}
\label{thm4}
Assume that $m\geq n_u$ experiments performed on the system \eqref{system} are conducted using multisines of the form \eqref{input}, designed under Assumptions \ref{assumption1} and \ref{assumption2}, and that the output vector of each experiment is in a stationary regime. Furthermore, assume that the output noise $\mathbf{v}(kh)$ is a Gaussian white noise signal of covariance $\bm{\Sigma}$. Then, the least-squares estimator $\hat{\bm{\mathcal{G}}}_{\textnormal{MS}}$ of $\bm{\mathcal{G}}_{0}$ is a sufficient statistic for $\bm{\theta}$.
\end{theorem}

\textit{Proof.} A consequence of the Fisher–Neyman factorization theorem \cite[Theorem 6.2.6]{casella2002statistical} is that if the data can be expressed as $\mathbf{y}^\top(kh) = \bm{\psi}^\hop(kh)\mathbf{g}(\bm{\theta}) + \mathbf{v}^\top(kh)$, where $\mathbf{v}^\top(kh)$ is i.i.d. Gaussian noise, $\bm{\psi}(kh)$ is a deterministic matrix, and $\mathbf{g}$ depends only on $\bm{\theta}$, then a sufficient statistic for $\bm{\theta}$ is given~by
\begin{equation}
    \hat{\mathbf{g}}(\mathbf{y}) = \left[\sum_{k=1}^N \bm{\psi}(kh)\bm{\psi}^\hop(kh) \right]^{-1} \left[\sum_{k=1}^N \bm{\psi}(kh)\mathbf{y}^\top(kh)\right]. \notag 
\end{equation}
In this context, we follow similar lines to \eqref{xkh} to find that $\mathbf{x}_i(kh,\bm{\theta})=\bm{\mathcal{G}}^\hop(\bm{\theta}) \bm{\zeta}_i(kh)$, which leads to the model
\begin{equation}
    \bm{\mathcal{Y}}^\top(kh) = \bm{\mathcal{Z}}^\hop(kh) \bm{\mathcal{G}}(\bm{\theta}) + \bm{\mathcal{V}}^\top(kh), \notag 
\end{equation}
where $\bm{\mathcal{Z}}(kh)$ is given by \eqref{zetamatrix}, and $\bm{\mathcal{V}}(kh)$ is the matrix formed by stacking $\mathbf{v}_i(kh)$, $i=1,\dots,m$. Exploiting this description together with the identities $ \bm{\mathcal{U}}(kh)=(\tilde{\bm{\Gamma}}\otimes \mathbf{I}_{n_u}) \bm{\mathcal{Z}}(kh)$ and $ (\tilde{\bm{\Gamma}}^\hop\otimes \mathbf{I}_{n_u}) \hat{\bm{\mathcal{H}}}_{\textnormal{MS}}=\hat{\bm{\mathcal{G}}}_{\textnormal{MS}}$ leads to writing a sufficient statistic for $\bm{\theta}$ as
\begin{align}
    \hat{\mathbf{g}}(\mathbf{y})  &= \left[\sum_{k=1}^N \bm{\mathcal{Z}}(kh)\bm{\mathcal{Z}}^\hop(kh) \right]^{-1} \sum_{k=1}^N \bm{\mathcal{Z}}(kh)\bm{\mathcal{Y}}^\top(kh) \notag \\
    &= (\tilde{\bm{\Gamma}}^{\hop}\hspace{-0.09cm}\otimes\hspace{-0.04cm} \mathbf{I}_{n_u} \hspace{-0.03cm})\hspace{-0.09cm}\left[\sum_{k=1}^N \bm{\mathcal{U}}\hspace{-0.02cm}(kh)\bm{\mathcal{U}}^{ \hspace{-0.04cm}\top}\hspace{-0.05cm}(kh)  \hspace{-0.02cm}\right]^{\hspace{-0.06cm}-\hspace{-0.03cm}1}\hspace{-0.12cm} \sum_{k=1}^N \hspace{-0.05cm} \bm{\mathcal{U}}\hspace{-0.02cm}(kh)\bm{\mathcal{Y}}^{\hspace{-0.04cm}\top}\hspace{-0.05cm}(kh) \hspace{-0.02cm} \notag \\
    &= (\tilde{\bm{\Gamma}}^\hop\otimes \mathbf{I}_{n_u}) \hat{\bm{\mathcal{H}}}_{\textnormal{MS}} \notag \\
    &= \hat{\bm{\mathcal{G}}}_{\textnormal{MS}}, \notag 
\end{align}
which shows what we aimed to prove. \hfill $\square$

In other words, when the noise is Gaussian-distributed, the information that the input and output data $\{\mathbf{u}_i(kh),\mathbf{y}_i(kh)\}_{k=1}^N$ for $i=1,\dots,m$ contains about $\bm{\theta}$ is completely captured by the least-squares estimator of the frequency response at each positive and negative input frequency. This result enables a link between the direct parametric method using the time-domain data and the two-step approach based on the frequency response estimate obtained in Section \ref{sec:ls}, as presented in the following corollary.

\begin{corollary}
\label{cor32}
Assume that the additive measurement noises $\{\mathbf{v}(kh)\}_{k=1}^N$ are Gaussian and white, with known covariance $\bm{\Sigma}$, the inputs in \eqref{input} satisfy Assumptions \ref{assumption1} and \ref{assumption2}, and that the outputs $\mathbf{y}_i(kh)$ are measured in a stationary regime. Then, the maximum likelihood  estimator of $\bm{\theta}$ within the compact set $\mathcal{D}$, based on the data $\{\mathbf{u}_i(kh),\mathbf{y}_i(kh)\}_{k=1}^N$ for $i=1,\dots,m$ is
\begin{equation}
\label{costipem}
\hat{\bm{\theta}} = \underset{\bm{\theta}\in \mathcal{D}}{\arg\min}  \left\|\textnormal{vec}\{\hat{\bm{\mathcal{G}}}_{\textnormal{MS}}\hspace{-0.04cm}-\hspace{-0.04cm}\bm{\mathcal{G}}(\bm{\theta})\hspace{-0.02cm}\}\right\|^2_{\textnormal{Cov}\{\textnormal{vec}\{\hat{\bm{\mathcal{G}}}_{\textnormal{MS}}\}\}^{-1}},
\end{equation}
where $\textnormal{Cov}\{\textnormal{vec}\{\hat{\bm{\mathcal{G}}}_{\textnormal{MS}}\}\}$ and $\bm{\mathcal{G}}(\bm{\theta})$ are given by \eqref{covariancegf} and \eqref{Gftheta} respectively. This estimator also minimizes the prediction error within the set $\mathcal{D}$
\begin{equation}
\label{costpem}
V(\bm{\theta}):= \sum_{i=1}^m \sum_{k=1}^N \big\|\mathbf{y}_i(kh)-\mathbf{y}_i(kh,\bm{\theta}) \big\|^2_{\bm{\Sigma}^{-1}},
\end{equation}
where
\begin{align}
\mathbf{y}_i(kh, \bm{\theta}) &= \mathbf{G}(0,\bm{\theta})\mathbf{a}_{0,i} \notag \\
&+ \sum_{\ell=1}^M \textnormal{Re}\left\{ \mathbf{G}(\mathrm{i}\omega_\ell,\bm{\theta})\big(\mathbf{a}_\ell \odot e^{\mathrm{i}\bm{\phi}_{\ell,i}}\big)e^{\mathrm{i}\omega_\ell kh} \right\}. \notag
\end{align}
Additionally, if the no spectral leakage assumption in Assumption \ref{assumption3} also holds,
\begin{align}
    \hat{\bm{\theta}} &=\underset{\bm{\theta}\in \mathcal{D}}{\arg\min} \textnormal{ tr}\bigg\{\bm{\Sigma}^{-1}\Big[\tilde{\mathbf{G}}(0,\bm{\theta}) \mathbf{A}_0^{\top}\mathbf{A}_0 \tilde{\mathbf{G}}^\top(0,\bm{\theta}) \notag \\
    \label{explicitipem}
    &+2\sum_{\ell=1}^M \textnormal{Re}\big\{\tilde{\mathbf{G}}(\mathrm{i}\omega_\ell,\bm{\theta}) \overline{\mathbf{A}_\ell^{\hop} \mathbf{A}_\ell} \tilde{\mathbf{G}}^\hop(\mathrm{i}\omega_\ell,\bm{\theta})\big\} \Big]\bigg\},
\end{align}
where $\tilde{\mathbf{G}}(\mathrm{i}\omega_\ell,\bm{\theta})=\hat{\mathbf{G}}(\mathrm{i}\omega_\ell)-\mathbf{G}(\mathrm{i}\omega_\ell,\bm{\theta}), \ell =0,1,\dots,M$.
\end{corollary}

\textit{Proof.} Since $\bm{\Sigma}$ is known, the log-likelihood function for $\mathbf{y}_i(kh)$, $k=1,\dots,N$, $i=1,\dots,m$ is proportional to $V(\bm{\theta})$. The statistical sufficiency of $\hat{\bm{\mathcal{G}}}_{\textnormal{MS}}$ implies that the log-likelihood is also proportional to the cost in  \eqref{costipem}. Thus, since these expressions only differ by a constant term, they share the same global optima. 

For reaching \eqref{explicitipem}, under Assumption \ref{assumption3} we have
\begin{equation}
    \textnormal{Cov}\{\textnormal{vec}\{\hat{\bm{\mathcal{G}}}_{\textnormal{MS}}\}\}^{-1} = N\bm{\Sigma}^{-1} \otimes \bm{\mathcal{A}}, \notag 
\end{equation}
where
\begin{equation}
\label{mathcalA}
\bm{\mathcal{A}}\hspace{-0.04cm}=\hspace{-0.07cm}\textnormal{blkdiag}(\hspace{-0.02cm}\mathbf{A}_0^{\hspace{-0.03cm}\top}\hspace{-0.03cm}\mathbf{A}_0, \hspace{-0.02cm}\overline{\mathbf{A}_1^{\hspace{-0.02cm}\hop} \hspace{-0.03cm}\mathbf{A}_1}\hspace{-0.01cm},\hspace{-0.02cm}\mathbf{A}_1^{\hspace{-0.02cm}\hop}\hspace{-0.03cm}\mathbf{A}_1,\hspace{-0.01cm}\dots\hspace{-0.02cm}, \overline{\mathbf{A}_{\hspace{-0.02cm}M}^\hop \hspace{-0.02cm}\mathbf{A}_M}\hspace{-0.02cm},\hspace{-0.01cm} \mathbf{A}_{\hspace{-0.02cm}M}^\hop \hspace{-0.02cm}\mathbf{A}_M\hspace{-0.02cm}).
\end{equation}
The vectorization identities $\textnormal{vec}\{\mathbf{ABC}\} = (\mathbf{C}^\top\otimes \mathbf{A})\textnormal{vec}\{\mathbf{B}\}$ and $\textnormal{tr}\{\mathbf{A}^\hop \mathbf{B}\} = \textnormal{vec}\{\mathbf{A}\}^\hop\textnormal{vec}\{\mathbf{B}\}$, when applied to \eqref{costipem}, yield
\begin{equation}
    \hat{\bm{\theta}}\hspace{-0.03cm}=\hspace{-0.03cm} \underset{\bm{\theta}\in \mathcal{D}}{\arg\min} \textnormal{ tr}\hspace{-0.04cm}\left\{\bm{\Sigma}^{-1}\big(\hat{\bm{\mathcal{G}}}_{\textnormal{MS}}-\bm{\mathcal{G}}(\bm{\theta}) \big)^\hop \bm{\mathcal{A}}\big(\hat{\bm{\mathcal{G}}}_{\textnormal{MS}}-\bm{\mathcal{G}}(\bm{\theta}) \big)\right\}. \notag 
\end{equation}
Expressing $\hat{\bm{\mathcal{G}}}_{\textnormal{MS}}$, $\bm{\mathcal{G}}(\bm{\theta})$ and $\bm{\mathcal{A}}$ in terms of $\mathbf{G}(\mathrm{i}\omega_\ell,\bm{\theta})$, $\mathbf{G}(\mathrm{i}\omega_\ell)$ and $\mathbf{A}_\ell$ respectively, gives \eqref{explicitipem}.
\hfill $\square$

\begin{remark}
    Corollary \ref{cor32} can also be interpreted through the indirect PEM framework \cite{soderstrom1991indirect}, which first fits an overparameterized model, here the FRF at $\omega=0,\pm\omega_1,\dots,\pm\omega_M$, and then projects the estimates onto a smaller parametric model structure, $\mathbf{G}(p,\bm{\theta})$, see Figure \ref{fig_ipem}. Since the residual in \eqref{HMS} is affine in the estimated parameters and \eqref{costipem} uses the exact covariance matrix \eqref{covariancegf}, the extended invariance principle \cite{stoica1989reparametrization} applies, yielding an alternative proof of the finite sample equivalence between the direct and indirect PEM formulations.
    \begin{figure}
	\centering{
		\includegraphics[width=0.47\textwidth]{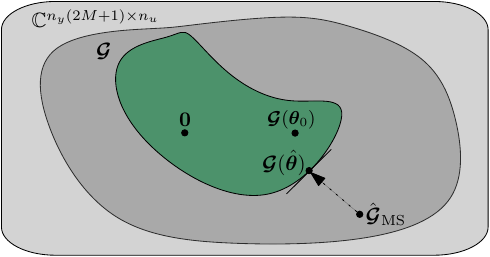}
		\caption{Interpretation of the indirect prediction error method for system identification under multisine excitation. The FRF estimate $\hat{\bm{\mathcal{G}}}_{\textnormal{MS}}$, lying in the subspace $\bm{\mathcal{G}} \subset \mathbb{C}^{n_y(2M+1)\times n_u}$, is projected onto the subspace of parametric models, yielding $\hat{\bm{\theta}}$.}
		\label{fig_ipem}}
\vspace{-0.4cm}
\end{figure}
\end{remark}

Corollary \ref{cor32} implies that the time-domain prediction error method coincides with the optimally-weighted frequency-domain identification method when the frequency response function estimates are computed with the least-squares method in \eqref{ls}, and both maximize the likelihood function of the total output data, even if the measured outputs suffer from aliasing effects. This relationship is valid for any $N> 2M$ if the outputs are in a stationary regime. Hence, it is not merely asymptotic. If the noise covariance $\bm{\Sigma}$ is not known, then it can be estimated jointly via the logdet criterion in, e.g., \cite[Ch. 7]{soderstrom1988system}; the analysis of this case is not considered in this work.

Naturally, Corollary \ref{cor32} also has implications for SISO identification. Since the simplified refined instrumental variable method for continuous-time systems with multisine excitations in \cite{gonzalez2020consistent} provides at convergence a stationary point of the likelihood under mild conditions, it is a way to solve the minimization problem in \eqref{costipem} via iterations in the time domain. Conversely, Corollary \ref{cor32} shows that the frequency-domain methods in \cite{blom2010multivariable} and \cite{vanderHulst2025FrequencyStage} are equivalent to time-domain prediction error methods for stationary multisine data, provided the nonparametric FRF is estimated via \eqref{frf} and the frequency weighting matrices are set to the inverse covariance from Theorem \ref{thm3}.

The equivalence in Corollary \ref{cor32} provides additional insights into the maximum likelihood and prediction error method estimators for continuous-time system identification. In the following, we study two scenarios:
\begin{enumerate}[label=\arabic*.]
	\item The parametric model $\mathbf{G}(p,\bm{\theta})$ is underconstrained or fully constrained, i.e., $n_\theta \geq n_u n_y (2M+1)$;
    \item The parametric model $\mathbf{G}(p,\bm{\theta})$ is overconstrained, i.e., $n_\theta < n_u n_y (2M+1)$.
\end{enumerate}
For the under and fully constrained parametric model cases, we show in Corollary \ref{cor33} that the parameter vector $\hat{\bm{\theta}}$ that minimizes \eqref{costipem} and \eqref{costpem} can be computed explicitly.

\begin{corollary}
	\label{cor33}
	Assume that $m\geq n_u$ experiments performed on the system \eqref{system} are conducted using multisines of the form \eqref{input}, designed under Assumptions \ref{assumption1} and \ref{assumption2}, and that the output vector of each experiment is in a stationary regime. Moreover, assume that parameter dimension and sample size satisfy $n_\theta \geq n_u n_y(2M+1)$ and $N>2M$ respectively, and that there exists a parametric model $\mathbf{G}(p,\tilde{\bm{\theta}})$ with a real-valued parameter vector $\tilde{\bm{\theta}}\in \mathcal{D}\subseteq \mathbb{R}^{n_{\theta}}$ such that
	\begin{equation}
 \label{parametriccondition}
	\mathbf{G}(\hspace{-0.01cm}0,\hspace{-0.02cm}\tilde{\bm{\theta}})\hspace{-0.04cm} =\hspace{-0.04cm} \hat{\mathbf{G}}(0), \hspace{0.1cm} \mathbf{G}(\pm \mathrm{i}\omega_\ell,\hspace{-0.02cm}\tilde{\bm{\theta}})\hspace{-0.06cm} =\hspace{-0.05cm} \hat{\mathbf{G}}(\pm \mathrm{i}\omega_\ell), \hspace{0.1cm} \ell\hspace{-0.04cm}=\hspace{-0.04cm}1,2,\dots,M,
	\end{equation}
	where $\hat{\mathbf{G}}(0)$ and $\hat{\mathbf{G}}(\pm\mathrm{i}\omega_\ell)$, $\ell=1,\dots,M$, are obtained from~\eqref{frf}. Then, the following statements are true:
	\begin{enumerate}[label=\arabic*.]
		\item
		The vector $\tilde{\bm{\theta}}$ minimizes the costs in \eqref{costipem} and \eqref{costpem}.
		\item
		The frequency response of the parametric model, $\mathbf{G}(\mathrm{i}\omega ,\hat{\bm{\theta}})$, is unbiased at the frequencies $\omega=0,\pm \omega_1,\pm \omega_2,\dots,\pm \omega_M$, and the joint covariance of these estimates is given by \eqref{covariancegf}.
	\end{enumerate}
\end{corollary}

\textit{Proof.} Statement 1 follows from the fact that $\bm{\mathcal{G}}(\tilde{\bm{\theta}}) = \hat{\bm{\mathcal{G}}}_{\textnormal{MS}}$, which means that the cost function in \eqref{costipem} reaches zero, its minimum value. Note that $\tilde{\bm{\theta}}$ is not necessarily unique. Statement 2 follows directly from Statement 1 and Theorems~\ref{thm2} and \ref{thm3}. \hfill $\square$

\begin{remark}
    The transfer function $\mathbf{G}(p,\tilde{\bm{\theta}})$ with parameter vector $\tilde{\bm{\theta}}\in\mathbb{R}^{n_\theta}$ satisfying \eqref{parametriccondition} exists under mild conditions on the model structure. For example, if each entry of $\mathbf{G}_0(p)$ is independently parameterized by a static gain and a total of $2M$ poles and zeros, then it can be shown that $\mathbf{G}(p,\tilde{\bm{\theta}})$ satisfying \eqref{parametriccondition} exists for all vectors $\hat{\bm{\mathcal{G}}}_{\textnormal{MS}}\in \mathbb{C}^{n_u(2M+1)\times n_y}$ obtained from least squares, except at most for a set of FRF estimates of Lebesgue measure zero.
\end{remark}
    
That is, for a parameter vector dimension of $n_\theta\geq n_u n_y(2M+1)$, an estimator $\hat{\bm{\theta}}$ that minimizes the time-domain prediction error \eqref{costpem} for finite samples $N>2M$ can be constructed by first forming the least-squares estimator of the frequency response at the input frequencies via \eqref{ls} and~\eqref{frf}, and later performing a matrix function interpolation. Note that the parameter vector must provide sufficient flexibility for \eqref{parametriccondition} to be feasible, although the parametric model is not restricted to independently parameterized scalar transfer functions.

For the overconstrained parametric model case, the optimization problem in \eqref{costipem} admits no closed-form solution and is typically addressed as a nonconvex problem using gradient-based iterations \cite{ljung1998system}. Nonetheless, it is possible to determine a concentration bound for the FRF of the parametric model evaluated at the input frequencies, as well as for the parametric estimator, relying on the finite-time statistical properties of $\hat{\bm{\mathcal{G}}}_{\textnormal{MS}}$ explored in Theorems \ref{thm2} and \ref{thm3} of this paper. For the parametric estimator concentration bound, it is assumed that a true parameter vector $\bm{\theta}_0 \in \mathcal{D}$ exists that generates the data and that no undermodeling occurs. Only a local probability bound is obtained in this case, since the proof relies on the mapping $\bm{\theta} \mapsto \bm{\mathcal{G}}(\bm{\theta})$ being bi-Lipschitz, a property that can only be guaranteed locally without imposing constraints on the parametric model structure.

\begin{theorem}
\label{thmprobability}
    Assume that $m\geq n_u$ experiments performed on the system \eqref{system}, described by the true parameter vector $\bm{\theta}_0\in \mathcal{D}$, are conducted using multisines of the form \eqref{input}, designed under Assumptions \ref{assumption1} to \ref{assumption3}, and that the output vector of each experiment is in a stationary regime. Furthermore, assume that the output noise $\mathbf{v}(kh)$ is a Gaussian white noise signal of covariance $\bm{\Sigma}$, and that the no undermodeling occurs in the parametric model. Then, for $N>2M$, for all $\delta\in (0,1]$,
\begin{align}
    \mathbb{P}&\Bigg\{\|\bm{\mathcal{G}}(\hat{\bm{\theta}})-\bm{\mathcal{G}}(\bm{\theta}_0)\|_{\textnormal{F}}\geq 2\sqrt{\frac{\lambda_{\textnormal{max}}(\bm{\Sigma})}{N\min\limits_{\substack{0 \le \ell \le M}}\lambda_{\textnormal{min}}(\mathbf{A}_\ell^\hop \mathbf{A}_\ell)}}\notag \\
\label{probabilitybound1}
    &\times\left(\sqrt{2\log\left(\frac{2}{\delta}\right)}+\sqrt{n_u n_y(2M+1)}\right)\Bigg\}\leq \delta.
\end{align}
Additionally, let $\beta\in (0,1)$ and assume that the mapping $\bm{\theta} \mapsto\textnormal{vec}\{\bm{\mathcal{G}}(\bm{\theta})\}$ is continuously differentiable in $\mathcal{D}$. Assume that the Jacobian $\mathbf{J}(\bm{\theta}):= \partial \textnormal{vec}\{\bm{\mathcal{G}}(\bm{\theta})\}/ \partial \bm{\theta}^\top$ has full column rank for all $\bm{\theta}$ in an open ball $\mathcal{B}$ contained in the interior of $\mathcal{D}$, centered at $\bm{\theta}_0$, whose elements satisfy $\|\mathbf{J}(\bm{\theta})-\mathbf{J}(\bm{\theta}_0)\|\leq (1-\beta)\sigma_{\textnormal{min}}(\mathbf{J}(\bm{\theta}_0))$ for all $\bm{\theta}\in\mathcal{B}$. Then, if $\hat{\bm{\theta}}\in \mathcal{B}$, for $N>2M$ and all $\delta\in (0,1]$,
\begin{align}
    \mathbb{P}&\Bigg\{\|\hat{\bm{\theta}}\hspace{-0.03cm}-\hspace{-0.03cm}\bm{\theta}_0\|\hspace{-0.05cm}\geq \hspace{-0.05cm}\frac{2}{\beta\sigma_{\textnormal{min}}(\mathbf{J}(\bm{\theta}_0))}\hspace{-0.02cm} \sqrt{\frac{\lambda_{\textnormal{max}}(\bm{\Sigma})}{N\min\limits_{\substack{0 \le \ell \le M}}\lambda_{\textnormal{min}}(\mathbf{A}_\ell^\hop \mathbf{A}_\ell)}}\notag \\
\label{probabilitybound2}
    &\times\left(\sqrt{2\log\left(\frac{2}{\delta}\right)}+\sqrt{n_u n_y(2M+1)}\right)\Bigg\}\leq \delta.
\end{align}
\end{theorem}

\textit{Proof}: Denote $\bm{\Gamma}:=\textnormal{Cov}\{\textnormal{vec}\{\hat{\bm{\mathcal{G}}}_{\textnormal{MS}}\}\}$. Due to Theorems~\ref{thm2} and \ref{thm3} and the Gaussianity assumption on $\mathbf{v}(kh)$, the random variable $\mathbf{e}:=\bm{\Gamma}^{-1/2}\textnormal{vec}\{\hat{\bm{\mathcal{G}}}_{\textnormal{MS}}-\bm{\mathcal{G}}(\bm{\theta}_0)\}$ is distributed as a zero-mean complex multivariate Gaussian with covariance matrix $\mathbf{I}_{n_u n_y(2M+1)}$ and complementary covariance matrix \cite{schreier2010statistical} $\mathbf{C}=\mathbb{E}\{\mathbf{e}\mathbf{e}^\top\}$ given by
\begin{equation}
    \mathbf{C} = \mathbf{I}_{n_y} \otimes \left(\mathbf{Z}^{-1/2} \sum_{k=1}^N \hspace{-0.03cm}\bm{\mathcal{Z}}(kh) \bm{\mathcal{Z}}^\top(kh)  \mathbf{Z}^{-\top/2}\right), \notag
\end{equation}
which can be shown to satisfy the inequality $\mathbf{C}^\hop \mathbf{C}\preceq\mathbf{I}_{n_u n_y(2M+1)}$. Thus,
\begin{equation}
\hat{\bm{\theta}} = \underset{\bm{\theta}\in \mathcal{D}}{\arg\min}  \left\| \bm{\Gamma}^{-1/2}\textnormal{vec}\{\bm{\mathcal{G}}(\bm{\theta})\hspace{-0.04cm}-\hspace{-0.04cm}
\bm{\mathcal{G}}(\bm{\theta}_0)\hspace{-0.02cm}\}-\mathbf{e}\right\|^2.
\end{equation}
Evaluating the cost function above at $\bm{\theta}=\hat{\bm{\theta}}$ and $\bm{\theta}=\bm{\theta}_0$ yields
\begin{equation}
    \left\| \bm{\Gamma}^{-1/2}\textnormal{vec}\{\bm{\mathcal{G}}(\hat{\bm{\theta}})    \hspace{-0.04cm}-\hspace{-0.04cm}
    \bm{\mathcal{G}}(\bm{\theta}_0)\hspace{-0.02cm}\}-\mathbf{e}\right\|^2\leq \left\|\mathbf{e}\right\|^2, \notag 
\end{equation}
leading to the chain of inequalities
\begin{align}
    \left\|\textnormal{vec}\hspace{-0.025cm}\{\bm{\mathcal{G}}\hspace{-0.01cm}(\hspace{-0.01cm}\hat{\bm{\theta}}\hspace{-0.01cm})
    \hspace{-0.08cm}-\hspace{-0.065cm}\bm{\mathcal{G}}\hspace{-0.02cm}(\hspace{-0.01cm}\bm{\theta}_0\hspace{-0.01cm})\hspace{-0.025cm}\}\hspace{-0.025cm}\right\|_{\bm{\Gamma}^{\hspace{-0.02cm}-\hspace{-0.025cm}1}}^2 \hspace{-0.16cm}&\leq \hspace{-0.06cm} 2\textnormal{Re}\hspace{-0.06cm}\left\{\hspace{-0.05cm}\mathbf{e}^{\hspace{-0.02cm}\hop}\hspace{-0.01cm} \bm{\Gamma}^{-\hspace{-0.025cm}1\hspace{-0.01cm}/\hspace{-0.01cm}2}\textnormal{vec}\{\bm{\mathcal{G}}\hspace{-0.01cm}(\hspace{-0.01cm}\hat{\bm{\theta}}\hspace{-0.01cm})  
    \hspace{-0.08cm}-\hspace{-0.065cm}\bm{\mathcal{G}}\hspace{-0.01cm}(\hspace{-0.01cm}\bm{\theta}_0\hspace{-0.01cm})\hspace{-0.02cm}\} \hspace{-0.06cm}\right\}\notag \\
    &\leq 2 \|\mathbf{e}\| \left\|\textnormal{vec}\{\bm{\mathcal{G}}(\hat{\bm{\theta}})\hspace{-0.04cm}-\hspace{-0.04cm}\bm{\mathcal{G}}(\bm{\theta}_0)\}\right\|_{\bm{\Gamma}^{-1}}, \notag
\end{align}
i.e., $\|\textnormal{vec}\{\bm{\mathcal{G}}(\hat{\bm{\theta}})-\bm{\mathcal{G}}(\bm{\theta}_0)\}\|_{\bm{\Gamma}^{-1}}\leq 2\|\mathbf{e}\|$. From Assumption \ref{assumption3} we obtain $\bm{\Gamma}^{-1}=N\bm{\Sigma}^{-1}\otimes \bm{\mathcal{A}}$, with $\bm{\mathcal{A}}$ defined in \eqref{mathcalA}. Moreover, after applying eigenvalue properties of block diagonal matrices and Kronecker products, the minimum eigenvalue of $\bm{\Gamma}^{-1}$ is given by
\begin{align}
 \lambda_{\textnormal{min}}(\bm{\Gamma}^{-1}) &= N  \lambda_{\textnormal{min}}(\bm{\Sigma}^{-1})  \lambda_{\textnormal{min}}(\bm{\mathcal{A}})  \notag \\
 &= \frac{N}{\lambda_{\textnormal{max}}(\bm{\Sigma})}\min\limits_{\substack{0 \le \ell \le M}}\lambda_{\textnormal{min}}(\mathbf{A}_\ell^\hop \mathbf{A}_\ell). \notag 
\end{align}
Thus,
\begin{align}
    \|\bm{\mathcal{G}}(\hspace{-0.01cm}\hat{\bm{\theta}}\hspace{-0.01cm})\hspace{-0.06cm}-\hspace{-0.045cm}\bm{\mathcal{G}}(\hspace{-0.01cm}\bm{\theta}_0\hspace{-0.01cm})\|_{\textnormal{F}}\hspace{-0.05cm}&=\hspace{-0.06cm}\|\textnormal{vec}\{\bm{\mathcal{G}}(\hat{\bm{\theta}})\hspace{-0.04cm}-\hspace{-0.04cm}\bm{\mathcal{G}}(\bm{\theta}_0)\}\|\notag \\
    &\leq \hspace{-0.06cm} \frac{1}{\sqrt{\hspace{-0.02cm}\lambda_{\textnormal{min}}\hspace{-0.02cm}(\bm{\Gamma}^{-\hspace{-0.02cm}1}\hspace{-0.01cm})}} \hspace{-0.03cm}\left\|\textnormal{vec}\{\bm{\mathcal{G}}(\hspace{-0.01cm}\hat{\bm{\theta}}\hspace{-0.01cm})\hspace{-0.06cm}-\hspace{-0.05cm}\bm{\mathcal{G}}(\hspace{-0.01cm}\bm{\theta}_0\hspace{-0.01cm})\hspace{-0.02cm}\}\right\|_{\bm{\Gamma}^{-\hspace{-0.02cm}1}} \notag \\ 
    &\leq 2\sqrt{\frac{\lambda_{\textnormal{max}}(\bm{\Sigma})}{N\min\limits_{\substack{0 \le \ell \le M}}\lambda_{\textnormal{min}}(\mathbf{A}_\ell^\hop \mathbf{A}_\ell)}}\|\mathbf{e}\|.   \notag
\end{align}
The probability bound \eqref{probabilitybound1} follows from this inequality and Lemma \ref{lemmaconcentration} in Appendix \ref{app:concentration}, whereas the probability bound \eqref{probabilitybound2} follows from \eqref{probabilitybound1} and Lemma \ref{lemmadifferentiable} in Appendix \ref{app:differentiable}. \hfill $\square$ 

Theorem \ref{thmprobability} provides important insights on the dependence of the model structure, multisine amplitudes and number of frequency lines, noise gain, sample size, and system dimensions on parametric model learning. In particular, an increase in noise variance or a decrease in input amplitude or number of experiments leads to requiring more data samples to achieve the same confidence bounds, and systems of larger dimensions are typically more costly to identify. Furthermore, all quantities in the bounds in \eqref{probabilitybound1} and \eqref{probabilitybound2} can be computed explicitly beforehand except $\sigma_{\textnormal{min}}(\mathbf{J}(\bm{\theta}_0))$, which can be relaxed to $\min_{\bm{\theta}\in \mathcal{D}} \sigma_{\textnormal{min}}(\mathbf{J}(\bm{\theta}))$ if needed. 

The finite-time confidence region for $\bm{\theta}$ is conditioned on the event that the Jacobian $\mathbf{J}(\hat{\bm{\theta}})$ deviates from $\mathbf{J}(\bm{\theta}_0)$ by at most $(1-\beta)\sigma_{\textnormal{min}}(\mathbf{J}(\bm{\theta}))$ in operator norm, where $\beta\in (0,1)$ is a user-defined parameter. Choosing $\beta$ close to zero increases the probability of this conditioning event but leads to a looser confidence region bound. Although not explicitly derived in our result, a global probability result (i.e., for $\hat{\bm{\theta}}\in \mathcal{D}$) may be derived directly with our techniques in specific cases, such as for parametric models that are linear in~$\bm{\theta}$.

The main result in Theorem \ref{thmprobability} can directly be used to compute a finite-time variance bound of the parametric FRF estimates at the input frequencies, which is presented next.

\begin{corollary}
\label{corprobability}
Under the same experimental conditions as Theorem \ref{thmprobability} for obtaining \eqref{probabilitybound1}, for any $N>2M$,
\begin{align}
    \mathbb{E}&\left\{ \|\bm{\mathcal{G}}(\hat{\bm{\theta}})-\bm{\mathcal{G}}(\bm{\theta}_0)\|_{\textnormal{F}}^2 \right\}\notag \\
    &\leq \frac{4}{N}\frac{\lambda_{\textnormal{max}}(\bm{\Sigma})}{\min\limits_{\substack{0 \le \ell \le M}}\lambda_{\textnormal{min}}(\mathbf{A}_\ell^\hop \mathbf{A}_\ell)}\left(B^2 + 2\sqrt{2\pi}B+4\right), \notag
\end{align}
where $B:= \sqrt{n_u n_y(2M+1)}$.
\end{corollary}
\textit{Proof}. To ease notation, we define
\begin{align}
    A &= 2\sqrt{\frac{\lambda_{\textnormal{max}}(\bm{\Sigma})}{N\min\limits_{\substack{0 \le \ell \le M}}\lambda_{\textnormal{min}}(\mathbf{A}_\ell^\hop \mathbf{A}_\ell)}}. \notag 
\end{align}
From Theorem \ref{thmprobability}, $\|\bm{\mathcal{G}}(\hat{\bm{\theta}})-\bm{\mathcal{G}}(\bm{\theta}_0)\|_{\textnormal{F}}>\sqrt{\epsilon}$ occurs with probability at most $2\exp(-\frac{1}{2}(\frac{\sqrt{\epsilon}}{A}-B)^2)$, where for $\epsilon<(AB)^2$ the bound is vacuous. Thus, from the integral characterization of the expected value, we find
\begin{align}
    &\hspace{-0.1cm}\mathbb{E}\hspace{-0.04cm}\left\{ \hspace{-0.04cm}\|\bm{\mathcal{G}}(\hat{\bm{\theta}})\hspace{-0.06cm}-\hspace{-0.05cm}\bm{\mathcal{G}}(\bm{\theta}_0)\|_{\textnormal{F}}^2 \hspace{-0.02cm}\right\} \hspace{-0.07cm}=\hspace{-0.12cm}\int_{0}^\infty \hspace{-0.15cm}\mathbb{P}\hspace{-0.05cm}\left\{ \hspace{-0.03cm}\|\bm{\mathcal{G}}(\hat{\bm{\theta}})\hspace{-0.06cm}-\hspace{-0.05cm}\bm{\mathcal{G}}(\bm{\theta}_0)\|_{\textnormal{F}}\hspace{-0.04cm}>\hspace{-0.04cm}\sqrt{\epsilon}\right\}\hspace{-0.04cm}\textnormal{d}\epsilon \notag \\
    &\hspace{-0.14cm}\leq (AB)^2 + 2\int_{(AB)^2}^\infty \exp\bigg(-\frac{1}{2}\left[\frac{\sqrt{\epsilon}}{A}-B\right]^2 \bigg)\textnormal{d}\epsilon \notag \\
    &\hspace{-0.14cm}=(AB)^2+4A^2\left(1+B\sqrt{\frac{\pi}{2}}\right),\notag
\end{align}
which proves the statement. \hfill $\square$

\vspace{-0.2cm}
\section{Model parametrizations and examples}
\label{sec:parametrizations}
\vspace{-0.05cm}
The following two subsections explore the consequences of the main results in Section \ref{sec:parametric} in matrix fraction description parametrizations, and present a case study where closed-form computations and finite-time statistical properties are discussed in depth.

\vspace{-0.3cm}
\subsection{Model parametrization: Matrix fraction description}
\label{subsec:mfd}
A model parametrization that we study in further detail is the left matrix fraction description (LMFD)
\begin{equation}
\label{lmfd}
    \mathbf{G}(p,\bm{\theta}) = \mathbf{D}^{-1}(p,\bm{\theta})\mathbf{N}(p,\bm{\theta}),
\end{equation}
where we parameterize the $\mathbf{D}$ and $\mathbf{N}$ polynomial matrices as
\begin{align}
    \mathbf{D}(p,\bm{\theta}) &= \mathbf{I}_{n_y} + \mathbf{D}_1 p +\cdots, + \mathbf{D}_{n_D}p^{n_D}, \notag \\
    \mathbf{N}(p,\bm{\theta}) &= \mathbf{N}_0 + \mathbf{N}_1 p +\cdots, + \mathbf{N}_{n_N}p^{n_N}, \notag
\end{align}
with $\mathbf{D}_i\in\mathbb{R}^{n_y\times n_y}$, and $\mathbf{N}_j\in \mathbb{R}^{n_y\times n_u}$. The parameter vector is described in terms of $\mathbf{D}_i$ and $\mathbf{N}_j$ as follows:
\begin{equation}
\label{thetas}
    \bm{\Theta}\hspace{-0.05cm}=\hspace{-0.05cm}[\mathbf{D}_1, \dots, \mathbf{D}_{n_D}\hspace{-0.02cm},\mathbf{N}_0,\dots, \mathbf{N}_{n_N}]^{\hspace{-0.02cm}\top}\hspace{-0.04cm}, \hspace{0.08cm} \bm{\theta}\hspace{-0.05cm}=\hspace{-0.05cm}\textnormal{vec}\{\bm{\Theta}\}.
\end{equation}
For the underconstrained and fully constrained model cases described in Section \ref{sec:parametric}, i.e., $n_D n_y + n_N n_u\geq 2Mn_u$, Corollary \ref{cor33} shows that the maximum likelihood estimator for $\bm{\theta}$ can be obtained through matrix rational interpolation \cite{cuyt1987nonlinear}, which for this case has an explicit closed-form solution. This solution provides the exact prediction error method and maximum likelihood estimator for finite sample size $N$ and any number of experiments $m\geq n_u$, and is presented next.
\begin{theorem}
\label{thm6}
   Assume that $m\geq n_u$ experiments performed on the system \eqref{system} are conducted using multisines of the form \eqref{input}, designed under Assumptions \ref{assumption1} and \ref{assumption2}, and that the output vector of each experiment is in a stationary regime. Moreover, consider the model described by the LMFD in \eqref{lmfd}, and assume that $n_D n_y + n_N n_u\geq 2Mn_u$ and $N>2M$. Let
\begin{equation}
\label{matrixj}
    \mathbf{J} \hspace{-0.05cm}= \hspace{-0.1cm}\begin{bmatrix}
        -\boldsymbol{\Omega}^1 \hspace{-0.01cm}\hat{\bm{\mathcal{G}}}_{\textnormal{MS}}, \dots\hspace{-0.02cm}, \hspace{-0.02cm}-\hspace{-0.02cm}\boldsymbol{\Omega}^{n_D}\hspace{-0.02cm} \hat{\bm{\mathcal{G}}}_{\textnormal{MS}}, \hspace{-0.02cm}\boldsymbol{\Omega}^0 \bm{\mathcal{I}},\dots, \hspace{-0.02cm}\boldsymbol{\Omega}^{n_N} \hspace{-0.03cm}\bm{\mathcal{I}}
    \end{bmatrix}\hspace{-0.02cm},
\end{equation}
where $\bm{\mathcal{I}}=[\mathbf{I}_{n_u},\dots,\mathbf{I}_{n_u}]^\top\in\mathbb{R}^{n_u(2M+1)\times n_u}$, and $\boldsymbol{\Omega} = \textnormal{diag}(0,-\mathrm{i}\omega_1,\mathrm{i}\omega_1,\dots,-\mathrm{i}\omega_M,\mathrm{i}\omega_M)\otimes \mathbf{I}_{n_u}\in \mathbb{C}^{n_u(2M+1)\times n_u(2M+1)}$, with $\boldsymbol{\Omega}^0=\mathbf{I}_{n_u(2M+1)}$.   

   Then, provided the matrix $\mathbf{J}$ has full row rank, all the parameters that jointly minimize the cost functions in \eqref{costipem} and \eqref{costpem} are given by
\begin{equation}
\label{thetahat}
    \hat{\bm{\Theta}}= \mathbf{J}^\dagger \hat{\bm{\mathcal{G}}}_{\textnormal{MS}} + \mathbf{K}\mathbf{L},    
\end{equation}
where $\mathbf{K}$ is a matrix whose $n_D n_y + (n_N-2M)n_u$ columns form a real basis of $\textnormal{ker}(\mathbf{J})$, and $\mathbf{L}\in \mathbb{R}^{[n_D n_y + (n_N\hspace{-0.03cm}-\hspace{-0.02cm}2M)n_u]\times n_y}$ is a matrix of free parameters.
\end{theorem}

\textit{Proof.} It follows from Corollary \ref{cor33} that, for $n_D n_y + n_N n_u\geq 2Mn_u$, the vector of parameters must satisfy $\mathbf{D}^{-1}(0,\hat{\bm{\theta}})\mathbf{N}(0,\hat{\bm{\theta}})=\hat{\mathbf{G}}(0)$ and  $\mathbf{D}^{-1}(\pm \mathrm{i}\omega_\ell,\hat{\bm{\theta}})\mathbf{N}(\pm \mathrm{i}\omega_\ell,\hat{\bm{\theta}})=\hat{\mathbf{G}}(\pm\mathrm{i}\omega_\ell)$, $\ell=1,\dots,M$. Denote the denominator and numerator polynomial matrix coefficients composing the PEM parameter estimate $\hat{\bm{\theta}}$ as  $\hat{\mathbf{D}}_i$ and $\hat{\mathbf{N}}_j$, $i=1,\dots,n_D$, $j=0,\dots,n_N$. After some matrix manipulations, the polynomial matrices $\mathbf{N}(p,\hat{\bm{\theta}})$ and $\mathbf{D}(p,\hat{\bm{\theta}})$ can be shown to satisfy, respectively,
\begin{equation}
\label{equate1}
    \begin{bmatrix}
        \mathbf{N}^\top(0,\hat{\bm{\theta}}) \\
        \mathbf{N}^\top(-\mathrm{i}\omega_1,\hat{\bm{\theta}})\\
        \mathbf{N}^\top(\mathrm{i}\omega_1,\hat{\bm{\theta}}) \\
        \vdots \\
        \mathbf{N}^\top(-\mathrm{i}\omega_M,\hat{\bm{\theta}})\\
        \mathbf{N}^\top(\mathrm{i}\omega_M,\hat{\bm{\theta}})
    \end{bmatrix} = \sum_{k=0}^{n_N} \boldsymbol{\Omega}^k \bm{\mathcal{I}} \hat{\mathbf{N}}_k^\top,
\end{equation}
and 
\begin{equation}
\label{equate2}
    \hspace{-0.2cm}\begin{bmatrix}
        \hat{\mathbf{G}}^\top(0)\mathbf{D}^\top(0,\hat{\bm{\theta}}) \\
        \hat{\mathbf{G}}^{\hspace{-0.02cm}\top}\hspace{-0.05cm}(-\mathrm{i}\omega_1)\mathbf{D}^\top(-\mathrm{i}\omega_1,\hat{\bm{\theta}})\\
        \hat{\mathbf{G}}^\top(\mathrm{i}\omega_1)\mathbf{D}^\top(\mathrm{i}\omega_1,\hat{\bm{\theta}}) \\
        \vdots \\
        \hat{\mathbf{G}}^{\hspace{-0.03cm}\top}\hspace{-0.06cm}(\hspace{-0.02cm}-\mathrm{i}\omega_{\hspace{-0.02cm}M}\hspace{-0.01cm})\mathbf{D}^{\hspace{-0.03cm}\top}\hspace{-0.05cm}(\hspace{-0.02cm}-\mathrm{i}\omega_{\hspace{-0.02cm}M},\hspace{-0.02cm}\hat{\bm{\theta}})\\
        \hat{\mathbf{G}}^{\hspace{-0.02cm}\top}\hspace{-0.05cm}(\mathrm{i}\omega_M)\mathbf{D}^\top(\mathrm{i}\omega_M,\hat{\bm{\theta}})
    \end{bmatrix} \hspace{-0.11cm}=\hspace{-0.03cm} \hat{\bm{\mathcal{G}}}_{\hspace{-0.02cm}\textnormal{MS}} \hspace{-0.03cm}+\hspace{-0.05cm} \sum_{k=1}^{n_D} \hspace{-0.03cm}\boldsymbol{\Omega}^k \hat{\bm{\mathcal{G}}}_{\hspace{-0.02cm}\textnormal{MS}} \hat{\mathbf{D}}_k^{\hspace{-0.02cm}\top}, \hspace{-0.05cm}
\end{equation}
where $\bm{\mathcal{I}}$ and $\boldsymbol{\Omega}$ are defined in the theorem statement. Thus, by equating \eqref{equate1} with \eqref{equate2}, the interpolation conditions can be written as $\mathbf{J}\hat{\bm{\Theta}}=\hat{\bm{\mathcal{G}}}_{\textnormal{MS}}$, with $\mathbf{J}$ as defined in \eqref{matrixj}. Such underdetermined linear matrix equation is solved explicitly by \eqref{thetahat}, see \cite[Chapter 3]{ben2003generalized} for more details.  \hfill $\square$

An important observation from Theorem \ref{thm6} is that statistically optimal parametric estimators can be computed without iterative procedures or solving non-convex optimization problems, even if the model structure is nonlinear in the parameters. In the underconstrained model case, the flexibility in choosing the matrix $\mathbf{L}$ in \eqref{thetahat} allows the direct fitting of models with lower-dimensional parametrizations. For instance, a modal or additive model \cite{vanderHulst2025FrequencyStage} of the form
\begin{equation}
\label{modalmodel}
    \mathbf{G}_m(p,\bm{\theta}) = \sum_{i=1}^n \frac{\bm{\phi}_{l,i}\bm{\phi}^\top_{r,i}}{a_{2,i}p^2 + a_{1,i}p+1}, 
\end{equation}
where $\bm{\phi}_{l,i}\in \mathbb{R}^{n_y}$, $\bm{\phi}_{r,i}\in \mathbb{R}^{n_u}$, and $a_{1,i}, a_{2,i}>0$, can be recovered as long as there exists a parameter vector $\tilde{\bm{\theta}}$ satisfying the interpolation conditions
\begin{equation}
\mathbf{G}_m\hspace{-0.02cm}(0,\tilde{\bm{\theta}}) \hspace{-0.04cm}=\hspace{-0.04cm} \hat{\mathbf{G}}(0), \mathbf{G}_m(\pm \mathrm{i}\omega_\ell,\hspace{-0.02cm}\tilde{\bm{\theta}}) \hspace{-0.04cm}=\hspace{-0.04cm} \hat{\mathbf{G}}(\pm \mathrm{i}\omega_\ell), \ell\hspace{-0.03cm}=\hspace{-0.03cm}1,\dots,M.    \notag 
\end{equation}
If $n_D=2n$ and $n_N\geq 2n-2$, Theorem \ref{thm6} guarantees the existence of a matrix $\mathbf{L}$ such that \eqref{thetahat} yields an LMFD model that can be decomposed exactly as \eqref{modalmodel}. Moreover, since the parametric model has an explicit dependency on $\hat{\bm{\mathcal{G}}}_{\textnormal{MS}}$, whose first and second order statistics are exactly known from Theorems \ref{thm2} and \ref{thm3}, tight confidence intervals and finite-sample distributions of the parameters can be obtained. 

Bear in mind that the solution of the normal equations $\mathbf{J}\hat{\bm{\Theta}}=\hat{\bm{\mathcal{G}}}_{\textnormal{MS}}$ can always be chosen as real-valued, since the conjugate parameter matrix estimate $\overline{\hat{\bm{\Theta}}}$ can be shown to satisfy the same normal equations. Moreover, the matrix $\mathbf{K}$ can always be taken to be real-valued because it can be computed from the real matrix formed by stacking the real and imaginary parts of the rows of $\mathbf{J}$. 

\begin{remark}
For the fully constrained model case, the closed-form solution provided in \eqref{thetahat} coincides with the one obtained from Levy's method for rational function estimation \cite{dyer2009least}. In fact, the interpolation equations formed by \eqref{equate1} and \eqref{equate2} correspond to the ones found in, e.g., Problem 2.3. of \cite{goodwin1977dynamic} for the SISO case. However, for the overconstrained model case, Levy's method solves the minimization problem
\begin{align}
\min_{\bm{\theta}\in\mathcal{D}} &\left\|\mathbf{D}(0,\bm{\theta})\hat{\mathbf{G}}(0) - \mathbf{N}(0,\bm{\theta})\right\|_{\mathbf{W_0}}^2 \notag \\
&+ 2\sum_{\ell=1}^M \left\|\mathbf{D}(\mathrm{i}\omega_\ell,\bm{\theta})\hat{\mathbf{G}}(\mathrm{i}\omega_\ell) - \mathbf{N}(\mathrm{i}\omega_\ell,\bm{\theta})\right\|_{\mathbf{W_\ell}}^2,   \notag
\end{align}
where $\mathbf{W}_\ell$, $\ell=0,1,\dots,M$ are fixed positive definite weighting matrices. This optimization problem does not have the same solution as \eqref{costipem} in general.
\end{remark}

\subsection{A case study in finite-sample behavior}
\label{subsec:finitetime}
As shown in Theorem \ref{thm6}, the least-squares estimator of the frequency response provides an approach to compute the maximum likelihood estimator in explicit form for the parametric model $\mathbf{G}(p,\bm{\theta})$, which is known to be a consistent and asymptotically efficient estimator in many scenarios \cite{ljung1998system}. To conclude this section, a case study is presented below, containing closed-form computations and finite-time statistical analyses on the maximum likelihood parametric estimate for a first-order system. 

Consider the true continuous-time system in transfer function form
\begin{equation}
\label{truesystem}
G_0(p)= \frac{b_{0,0}}{a_{1,0} p+1}, 
\end{equation}
where $a_{1,0}>0$ and $b_{0,0}\neq 0$. The output of $G_0(p)$ is corrupted by Gaussian white noise of variance $\sigma^2$. We are interested in obtaining an estimate of the system for three different scenarios: 1) the fully constrained model case with $G(p,\bm{\theta})=b_0/(a_1 p+1)$ and input $u(t)=\alpha_1 \cos(\omega_1 t-\pi/2)$ with $\alpha_1,\omega_1>0$; 2) the underconstrained model case with $G(p,\bm{\theta})=(b_1 p +b_0)/(a_1 p+1)$ and same input as 1); and 3) the overconstrained model case with $G(p,\bm{\theta})=b_0/(a_1 p+1)$ and input $u(t)=\alpha_0+\alpha_1 \cos(\omega_1 t-\pi/2)$ with $\alpha_0\neq 0, \alpha_1,\omega_1>0$. For simplicity, we assume that Assumption \ref{assumption3} holds, i.e., we consider $Nh$ to be a multiple of $2\pi/\omega_1$, so that the resulting frequency response estimates are uncorrelated according to Theorem \ref{thm3}.

\textit{1) Fully constrained model case}: By Corollary \ref{cor33}, the maximum likelihood estimator $\hat{\bm{\theta}} = [\hat{a}_1, \hat{b}_0]^\top$ must satisfy
\begin{equation}
\hat{G}(\mathrm{i}\omega_1) = \frac{\hat{b}_0}{\mathrm{i}\hat{a}_1 \omega_1 + 1} \quad \textnormal{and} \quad \hat{G}(-\mathrm{i}\omega_1) = \frac{\hat{b}_0}{-\mathrm{i}\hat{a}_1 \omega_1 + 1},\notag
\end{equation}
where $\hat{G}(\mathrm{i}\omega_1)$ and $\hat{G}(-\mathrm{i}\omega_1)$ are computed from the least-squares estimate of the FRF in \eqref{frf}. These equations form a system in $\hat{a}_1$ and $\hat{b}_0$ that can be solved explicitly, leading to
\begin{equation}
\label{a1b0}
\begin{bmatrix}
\hat{a}_1 \\
\hat{b}_0
\end{bmatrix} = \begin{bmatrix}
-\dfrac{\textnormal{Im}\{\hat{G}(\mathrm{i}\omega_1)\}}{\omega_1 \textnormal{Re}\{\hat{G}(\mathrm{i}\omega_1)\}} \\
\dfrac{|\hat{G}(\mathrm{i}\omega_1)|^2}{\textnormal{Re}\{\hat{G}(\mathrm{i}\omega_1)\}}
\end{bmatrix}.
\end{equation}
Let us focus on $\hat{a}_1$. This parameter is given by the quotient of two zero-mean random variables that are jointly Gaussian:
\begin{equation}
\begin{bmatrix}
\textnormal{Re}\{\hat{G}(\mathrm{i}\omega_1)\} \\
\textnormal{Im}\{\hat{G}(\mathrm{i}\omega_1)\}
\end{bmatrix} \sim \mathcal{N}\left(\begin{bmatrix}
\textnormal{Re}\{G_0(\mathrm{i}\omega_1)\} \\
\textnormal{Im}\{G_0(\mathrm{i}\omega_1)\}
\end{bmatrix}, \mathbf{P}\right), \notag
\end{equation}
where, after computing $\mathbf{Z}=(\alpha_1^2/4)\mathbf{I}_2$ in \eqref{z}, we find that the covariance matrix $\mathbf{P}$ is exactly given by
\begin{equation}
\mathbf{P}=\frac{\sigma^2}{4}\begin{bmatrix}
1 & 1 \\
-1 & 1
\end{bmatrix}\mathbf{Z}^{-1} \begin{bmatrix}
1 & -1 \\
1 & 1
\end{bmatrix} =\frac{2\sigma^2}{N \alpha_1^2}\mathbf{I}_2. \notag
\end{equation}
Thus, the distribution of $\hat{a}_1$ for finite $N$ is given by the ratio of independent normal random variables with non-zero means. The exact distribution for this scenario has been derived in \cite{hinkley1969ratio}, enabling the computation of tight confidence bounds. The distribution of $\hat{a}_1$ is heavy-tailed and does not have finite moments \cite{marsaglia2006ratios}; however, it can be approximated by a normal distribution under certain conditions involving the probability of observing negative (resp. positive) values of $\textnormal{Re}\{\hat{G}(\mathrm{i}\omega_1)\}$ if $\textnormal{Re}\{G_0(\mathrm{i}\omega_1)\}$ is positive (resp. negative). A precise statement concerning the normal approximation is given in \cite{diaz2013existence}, but for the sake of this example we consider a practical result. Using the criterion\footnote{Other criteria are possible, and they are usually related to how small the ratio between the standard deviation and the mean of the random variable of the denominator is. Such random variable in this case is $\omega_1 \textnormal{Re}\{\hat{G}(\mathrm{i}\omega_1)\}$.}stated in \cite{kuethe2000imaging}, whenever 
\begin{equation}
\label{criterion}
\frac{\sigma}{\alpha_1 |\textnormal{Re}\{G_0(\mathrm{i}\omega_1)\}|} \sqrt{\frac{2}{N}}\leq 0.1,
\end{equation}
the normal approximation is valid, and the normal distribution that approximates the distribution of $\hat{a}_1$ (see \cite[Eq. (9)]{diaz2013existence}) is 
\begin{align}
\hat{a}_1 \sim \mathcal{N}&\bigg(-\frac{\textnormal{Im}\{G_0(\mathrm{i}\omega_1)\}}{\omega_1 \textnormal{Re}\{G_0(\mathrm{i}\omega_1)\}}, \notag \\
&\frac{2\sigma^2}{N \alpha_1^2 \omega_1^2 \textnormal{Re}\{G_0(\mathrm{i}\omega_1)\}^2} \left[1+\frac{\textnormal{Im}\{G_0(\mathrm{i}\omega_1)\}^2}{\textnormal{Re}\{G_0(\mathrm{i}\omega_1)\}^2}\right] \bigg). \notag
\end{align}
Writing this distribution in terms of the true parameters $a_{1,0}$ and $b_{0,0}$ gives
\begin{equation}
\label{approxnormal}
\hat{a}_1 \sim \mathcal{N}\left(a_{1,0}, \frac{2\sigma^2 (1+a_{1,0}^2 \omega_1^2)^3}{N \alpha_1^2 \omega_1^2 b_{0,0}^2} \right).
\end{equation}
Four interesting aspects can be noted from the approximate mean and variance above: 
\begin{enumerate}
\item
For the normal approximation to be valid we require a balance between $\textnormal{Re}\{G_0(\mathrm{i}\omega_1)\}$ and $\sqrt{N}$ in \eqref{criterion}: If $\omega_1$ is too high, then $\textnormal{Re}\{G_0(\mathrm{i}\omega_1)\}$ will be small and more data points will be required for achieving a reasonable normal approximation. Also, the input signal-to-noise ratio $\alpha_1^2/\sigma^2$ plays a natural role in \eqref{criterion}: If such ratio is large, then less data points are needed for the approximation to be valid. 
\item
The mean of the approximate normal distribution of $\hat{a}_1$ equals the true parameter $a_{1,0}$. This approximation holds in finite time and can be empirically observed even with fewer than 100 samples. In Figure \ref{fig_example} we compare the empirical distribution (using $2\cdot 10^5$ Monte Carlo runs) with the normal approximation for two different sample sizes: $N=20$ and $N=60$. The following parameters are considered: $a_{1,0}=1$; $b_{0,0}=2$; $\sigma=0.8$; $\alpha_1=1$; $\omega_1 = 1/\sqrt{2}$. Despite the relatively low signal-to-noise ratio (approximately $5$ [\text{dB}]) and limited data, the approximation aligns well with the empirical distribution.

\begin{figure}
	\centering{
		\includegraphics[width=0.48\textwidth]{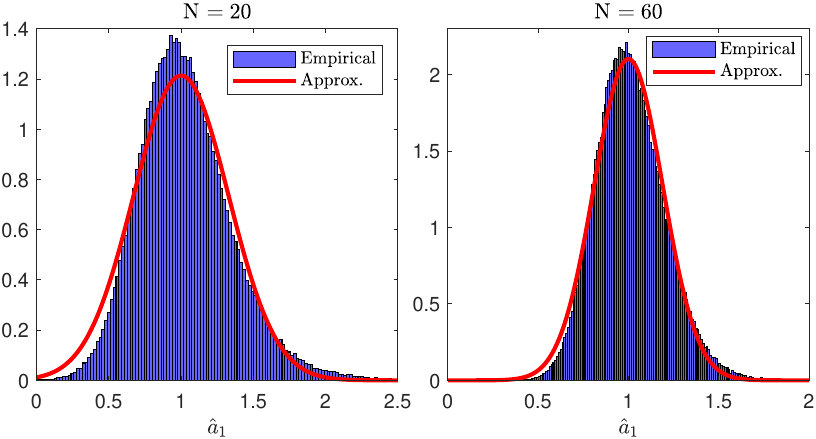}
        \vspace{-0.2cm}
		\caption{Empirical distribution of $\hat{a}_1$ (histogram) compared to its approximate normal distribution (red curve). Left: $N=20$; right: $N=60$. }
		\label{fig_example}}
        \vspace{-0.5cm}
\end{figure}

\item
The frequency that minimizes the (approximate) variance of the parameter $\hat{a}_1$ is $\omega_1 = 1/(\sqrt{2}a_{1,0})$, which corresponds to a $1.18$ $[\textnormal{dB}]$ attenuation from the steady state gain. Notably, the cutoff frequency $\omega_1 = 1/a_{1,0}$ does not minimize the approximate variance. The optimal approximate variance is given by
\begin{equation}
\textnormal{Var}\{\hat{a}_1\}\approx \frac{27 \sigma^2 a_{1,0}^2}{2N \alpha_1^2 b_{0,0}^2}. \notag
\end{equation}
\item
The approximate variance in \eqref{approxnormal} coincides with the asymptotic variance obtained through standard methods (see Appendix \ref{sec:standardascov} for the details). However, our expression was obtained through a finite-time analysis, leading to greater insight about when the estimates can be considered to have reached their asymptotic distribution.
\end{enumerate}
\textit{2) Underconstrained model case}: Now consider the model $G(p,\bm{\theta}) = (b_1p+b_0)/(a_1p+1)$, with $\bm{\theta} = [a_1,b_0,b_1]^\top$. The maximum likelihood estimator is non-unique in this scenario, and thus Theorem \ref{thm6} provides the set of all estimators that minimize the prediction error. 

The kernel of $\mathbf{J}$ in \eqref{matrixj} is described by the basis vector
\begin{equation}
    \mathbf{K} = \begin{bmatrix}
        \dfrac{1}{\textnormal{Re}\{\hat{G}(\mathrm{i}\omega_1)\}}, &
        -\omega_1\dfrac{\textnormal{Im}\{\hat{G}(\mathrm{i}\omega_1)\}}{\textnormal{Re}\{\hat{G}(\mathrm{i}\omega_1)\}}, &
        1
    \end{bmatrix}^\top.
\end{equation}
Since it is known from the fully constrained model case that one solution of the rational interpolation problem $\mathbf{J}\hat{\bm{\Theta}}=\hat{\bm{\mathcal{G}}}_{\textnormal{MS}}$ is given by $\hat{b}_1=0$ and $\hat{a}_1,\hat{b}_0$ given by \eqref{a1b0}, then it follows from Theorem \ref{thm6} that all maximum likelihood estimators for $\bm{\theta}$ can be written as
\begin{equation}
    \begin{bmatrix}
        \hat{a}_1 \hspace{-0.02cm}\\
        \hat{b}_0 \hspace{-0.02cm}\\
        \hat{b}_1\hspace{-0.02cm}
    \end{bmatrix} \hspace{-0.14cm}= \hspace{-0.14cm}\begin{bmatrix}
\dfrac{-\textnormal{Im}\{\hat{G}(\mathrm{i}\omega_1)\}}{\omega_1 \textnormal{Re}\{\hat{G}(\mathrm{i}\omega_1)\}} \\
\dfrac{|\hat{G}(\mathrm{i}\omega_1)|^2}{\textnormal{Re}\{\hat{G}(\mathrm{i}\omega_1)\}} \\
        0
    \end{bmatrix}\hspace{-0.1cm}+\hspace{-0.03cm}L\hspace{-0.05cm}\begin{bmatrix}
\dfrac{1}{\textnormal{Re}\{\hat{G}(\mathrm{i}\omega_1)\}} \\
 \dfrac{-\omega_1\textnormal{Im}\{\hat{G}(\mathrm{i}\omega_1)\}}{\textnormal{Re}\{\hat{G}(\mathrm{i}\omega_1)\}} \\
        1
    \end{bmatrix}\hspace{-0.07cm}, L\hspace{-0.03cm}\in\hspace{-0.03cm} \mathbb{R}. \notag 
\end{equation}
A more parsimonious model can be derived directly after choosing which parameters should be constrained to zero. As an example, a zero at zero frequency can be enforced by setting $\hat{b}_0=0$, leading to
\begin{equation}
\begin{bmatrix}
\hat{a}_1 \\
\hat{b}_1
\end{bmatrix} = \begin{bmatrix}
\dfrac{\textnormal{Re}\{\hat{G}(\mathrm{i}\omega_1)\}}{\omega_1 \textnormal{Im}\{\hat{G}(\mathrm{i}\omega_1)\}} \\
\dfrac{|\hat{G}(\mathrm{i}\omega_1)|^2}{\omega_1 \textnormal{Im}\{\hat{G}(\mathrm{i}\omega_1)\}}
\end{bmatrix}. \notag 
\end{equation}
Naturally, the resulting model is typically unstable if the true system is given by \eqref{truesystem}.

\textit{3) Overconstrained model case}: Since the maximum likelihood estimator has no closed-form solution for this case, a Monte Carlo simulation is presented to illustrate the confidence bounds in Theorem \ref{thmprobability}.

The system in \eqref{truesystem} with $a_1=1$ and $b_0=2$ is excited with $u(t)=1/2+\cos(2t-\pi/2)$, and the noise variance is set to $\sigma^2 = 0.25$. The signals are sampled at $h=\pi/10$. For an increasing sample size, empirical values for $B_1,B_2$ are computed such that each of the following events occur with probability at least 0.9:
\begin{align}
    E_1\hspace{-0.03cm}=\hspace{-0.03cm}&\left\{ \|\bm{\mathcal{G}}(\hat{\bm{\theta}})-\bm{\mathcal{G}}(\bm{\theta}_0)\|_{\textnormal{F}}<B_1\right\}, \notag \\
    E_2\hspace{-0.03cm}=\hspace{-0.03cm}&\left\{\|\hat{\bm{\theta}}\hspace{-0.04cm}-\hspace{-0.04cm}\bm{\theta}_0\|\hspace{-0.03cm}<\hspace{-0.03cm}B_2\hspace{0.05cm}\Big|\|\mathbf{J}(\hat{\bm{\theta}})\hspace{-0.07cm}-\hspace{-0.06cm}\mathbf{J}(\bm{\theta}_0)\|\hspace{-0.04cm}\leq\hspace{-0.04cm} 0.2\sigma_{\textnormal{min}}(\mathbf{J}(\bm{\theta}_0))\hspace{-0.02cm}\right\}\hspace{-0.05cm}. \notag 
\end{align}
These empirical values are compared with the theoretical bounds presented in Theorem \ref{thmprobability}. The sample size $N$ is chosen as a multiple of $10$, ranging from $N=20$ to $N=300$, and $2000$ Monte Carlo runs are performed for each $N$. Note that every sample size multiple of $10$ satisfies Assumption \ref{assumption3}. The time-domain simplified refined instrumental variable algorithm of \cite{gonzalez2020consistent} is used to compute the parametric estimator, where a relative tolerance of $10^{-15}$ is implemented for the termination rule of the refined instrumental variable iterations.
\begin{figure}
	\centering{
		\includegraphics[width=0.47\textwidth]{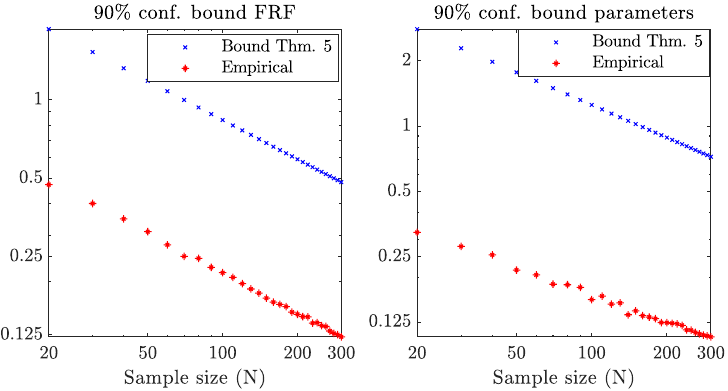}
		\vspace{-0.2cm}
        \caption{Empirical and theoretical $90\%$ confidence bounds for the FRF at the input frequency points (left), and for the vector $\hat{\bm{\theta}}$ describing the parametric model (right).}
		\label{fig_bound}}
        \vspace{-0.45cm}
\end{figure}

Figure \ref{fig_bound} compares empirical and theoretical bounds for the FRF and parameter confidence regions. The obtained confidence bounds exhibit the expected $1/N$ decay of the FRF and parameter errors in norm. In this example, the theoretical bound in \eqref{probabilitybound1} deviates by a factor of approximately $4$ from the empirical bounds for the FRF estimation, while a larger gap is observed for the parameter confidence bound due to the relaxation required for obtaining the bi-Lipschitz constant $\beta \sigma_{\textnormal{min}}(\mathbf{J}(\bm{\theta}_0))$ in Lemma \ref{lemmadifferentiable}.

\vspace{-0.2cm}
\section{Conclusions}
\label{sec:conclusions}
In this paper, we developed a framework to analyze finite sample properties of the least-squares method for continuous-time MIMO system identification. First, we derived a nonparametric FRF estimator from the least-squares estimator of the multisine-interpolated discrete-time equivalent response, and established its properties. These results revealed the effects of undersampling on FRF identification and lead to a consistency analysis for when the number of experiments grows.

The finite-time statistical properties of this estimator also enabled a formal link between frequency-domain parametric system identification and the time-domain prediction error method using concatenated input-output data. Such connection 1) clarifies the exact expression for optimal weighting in MIMO frequency-domain parametric identification; 2) provides a closed-form maximum likelihood estimator of a parametric model when the number of frequency lines does not exceed the number of parameters, and 3) enables finite-time confidence bounds for overconstrained models. A numerical case study was put forward to illustrate these findings.
\vspace{-0.2cm}

\appendices
\section{Norm concentration for Gaussian distributions}
\label{app:concentration}
\begin{lemma}
\label{lemmaconcentration}
    Consider the $m$-dimensional zero-mean complex Gaussian random variable $\mathbf{x}$ with covariance $\mathbf{I}_m$ and complementary covariance matrix $\mathbf{C}$, where $\mathbf{I}_m-\mathbf{C}^\hop \mathbf{C}$ is positive semidefinite. Then, for all $\delta\in (0,1]$,
\begin{equation}
\label{app1lemma}
    \mathbb{P}\left\{\|\mathbf{x}\| \geq \sqrt{2\log\left(\frac{2}{\delta}\right)}+\sqrt{m}\right\}\leq \delta.
\end{equation}    
\end{lemma}
\textit{Proof}. The complex random variable $\mathbf{x}$ can be characterized by the normal distribution of $\mathbf{y}=[\textnormal{Re}\{\mathbf{x}^\top\},\textnormal{Im}\{\mathbf{x}^\top\}]^\top$ as follows (see, e.g., \cite[p. 62]{schreier2010statistical}):
\begin{equation}
    \begin{bmatrix}
        \textnormal{Re}\{\mathbf{x}\} \\
        \textnormal{Im}\{\mathbf{x}\}
    \end{bmatrix} \sim \mathcal{N}\left(\mathbf{0},\frac{1}{2}\begin{bmatrix}
        \mathbf{I}_m+\textnormal{Re}\{\mathbf{C}\} & \textnormal{Im}\{\mathbf{C}\} \\
        \textnormal{Im}\{\mathbf{C}\} & \mathbf{I}_m-\textnormal{Re}\{\mathbf{C}\}
    \end{bmatrix}\right). \notag
\end{equation}
Using the Rayleigh quotient, the largest eigenvalue of $\textnormal{Cov}\{\mathbf{y}\}$ is bounded by
\begin{align}
    \lambda_{\textnormal{max}}(\textnormal{Cov}\{\mathbf{y}\})\hspace{-0.05cm}=\hspace{-0.05cm}\max_{\|\mathbf{z}\|=1} \frac{1\hspace{-0.05cm}+\hspace{-0.05cm}\textnormal{Re}\{\mathbf{z}^\top \mathbf{C}\mathbf{z}\}}{2} \leq \frac{1+\|\mathbf{C}\|}{2} \leq 1, \notag 
\end{align}
where the last inequality follows from $\mathbf{I}_m-\mathbf{C}^\hop \mathbf{C}$ being positive semidefinite. By considering the Cholesky decomposition $\mathbf{VV}^\top$ of the covariance matrix above, we note that the random variable $\mathbf{e}:=\mathbf{V}^{-1}\mathbf{y}$ is a standard multivariate Gaussian. The map $\mathbf{e}\mapsto \|\mathbf{Ve}\|=\|\mathbf{y}\|=\|\mathbf{x}\|$ is $1$-Lipschitz since for any vectors $\mathbf{e}_1,\mathbf{e}_2$ of appropriate dimensions,
\begin{align}
    \big| \|\mathbf{Ve}_1\|-\|\mathbf{Ve}_2\| \big| &\leq \|\mathbf{V}(\mathbf{e}_1-\mathbf{e}_2) \| \notag \\
    &\leq \sigma_{\textnormal{max}}(\mathbf{V})\|\mathbf{e}_1-\mathbf{e}_2\| \notag \\
    &\leq \|\mathbf{e}_1-\mathbf{e}_2\|. \notag 
\end{align}
Thus, Theorem 2.26 of \cite{wainwright2019high} gives the concentration bound
\begin{equation}
\label{app1proof}
    \mathbb{P}\left\{\|\mathbf{x}\| \geq t+\mathbb{E}\{\|\mathbf{x}\|\}\right\}\leq 2\exp (-t^2/2),
\end{equation}    
valid for all $t\geq 0$. By noting that $\mathbb{E}\{\|\mathbf{x}\|^2\}=\mathbb{E}\{\|\mathbf{Ve}\|^2\}=\textnormal{tr}\{\textnormal{Cov}\{\mathbf{y}\}\}=m$, Jensen's inequality gives $\mathbb{E}\{\|\mathbf{x}\|\}\leq \sqrt{m}$. Hence, $\{\mathbf{x}\colon \|\mathbf{x}\| \geq t+\sqrt{m}\}\subset \{\mathbf{x}\colon \|\mathbf{x}\| \geq t+\mathbb{E}\{\|\mathbf{x}\|\}$. Taking probability on both sides, introducing $\delta=2\exp (-t^2/2)$ and combining with \eqref{app1proof} gives \eqref{app1lemma}, concluding the proof. \hfill $\square$

\section{Local lower bound for differentiable maps}
\label{app:differentiable}
\begin{lemma}
\label{lemmadifferentiable}
Let the mapping $\bm{\theta}\to \textnormal{vec}\{\bm{\mathcal{G}}(\bm{\theta})\}$ be continuously differentiable on a convex and compact set $\mathcal{D}$, and let $\beta\in(0,1)$. Suppose that the Jacobian $\mathbf{J}(\bm{\theta})$ of $\textnormal{vec}\{\bm{\mathcal{G}}(\bm{\theta})\}$ has full column rank for all $\bm{\theta}\in\mathcal{D}$. Then, for all $\bm{\theta}_0$ there exists an open ball $\mathcal{B}\subset \mathcal{D}$ centered around $\bm{\theta}_0$ whose points satisfy $\|\mathbf{J}(\bm{\theta})-\mathbf{J}(\bm{\theta}_0)\|\leq (1-\beta)\sigma_{\textnormal{min}}(\mathbf{J}(\bm{\theta}_0))$. Additionally, for all $\bm{\theta}\in\mathcal{B}$, the following inequality holds:
\begin{equation}
\|\bm{\mathcal{G}}(\hspace{-0.01cm}\bm{\theta}\hspace{-0.01cm})\hspace{-0.06cm}-\hspace{-0.045cm}\bm{\mathcal{G}}(\hspace{-0.01cm}\bm{\theta}_0\hspace{-0.01cm})\|_{\textnormal{F}}\geq \beta \sigma_{\textnormal{min}}(\mathbf{J}(\bm{\theta}_0))\|\bm{\theta}\hspace{-0.06cm}-\hspace{-0.045cm}\bm{\theta}_0\|.
\end{equation}
\end{lemma}

\textit{Proof}. By assumption $\mathbf{J}(\bm{\theta})$ is continuous on the compact set $\mathcal{D}$, with $\bm{\theta}_0$ belonging to the interior of $\mathcal{D}$. Hence, there exists an open ball $\mathcal{B}\subset \mathcal{D}$ containing $\bm{\theta}_0$ such that $\|\mathbf{J}(\bm{\theta})-\mathbf{J}(\bm{\theta}_0)\|\leq (1-\beta)\sigma_{\textnormal{min}}(\mathbf{J}(\bm{\theta}_0))$ for all $\bm{\theta}\in \mathcal{B}$, where $\beta\in(0,1)$ and $\sigma_{\textnormal{min}}(\mathbf{J}(\bm{\theta}_0))>0$ due to the full column rank condition. Since $\bm{\theta}\mapsto \textnormal{vec}\{\bm{\mathcal{G}}(\bm{\theta})\}$ is continuously differentiable, the mean value theorem \cite[Theorem 4.2, p.341]{lang2012real} applies. Thus, for $\bm{\theta},\bm{\theta}_0\in \mathcal{B}$,
\begin{equation}
    \label{integral}
    \textnormal{vec}\{\bm{\mathcal{G}}(\hspace{-0.01cm}\bm{\theta}\hspace{-0.01cm})\hspace{-0.06cm}-\hspace{-0.045cm}\bm{\mathcal{G}}(\hspace{-0.01cm}\bm{\theta}_0\hspace{-0.01cm})\} \hspace{-0.08cm}=\hspace{-0.09cm} \int_0^1 \hspace{-0.07cm} \mathbf{J}(\bm{\theta}_0+t[\bm{\theta}\hspace{-0.04cm}-\hspace{-0.03cm}\bm{\theta}_0] )\textnormal{d}t (\bm{\theta}\hspace{-0.04cm}-\hspace{-0.03cm}\bm{\theta}_0).
\end{equation}
If we denote the integral in \eqref{integral} by $\mathcal{I}_{\bm{\theta}_0}(\bm{\theta}-\bm{\theta}_0)$, then 
\begin{align}
    \|\mathcal{I}_{\bm{\theta}_0}(\bm{\theta}\hspace{-0.04cm}-\hspace{-0.03cm}\bm{\theta}_0)\hspace{-0.05cm}-\hspace{-0.03cm}\mathbf{J}(\bm{\theta}_0) \| &\hspace{-0.03cm}\leq \hspace{-0.05cm}\int_0^1 \hspace{-0.05cm}\|\mathbf{J}(\bm{\theta}_0+t[\bm{\theta}\hspace{-0.06cm}-\hspace{-0.045cm}\bm{\theta}_0] )\hspace{-0.04cm}-\hspace{-0.03cm}\mathbf{J}(\bm{\theta}_0)\|\textnormal{d}t \notag \\
    &\leq \sup_{\bm{\xi}\in\mathcal{B}}\|\mathbf{J}(\bm{\xi})-\mathbf{J}(\bm{\theta}_0)\| \notag \\
    &\leq (1-\beta)\sigma_{\textnormal{min}}(\mathbf{J}(\bm{\theta}_0)). \notag 
\end{align}
Thus, Corollary 7.3.5 in \cite{Horn2012} yields the singular value inequality
\begin{align}
    \sigma_{\textnormal{min}}(\mathcal{I}_{\bm{\theta}_0}\hspace{-0.02cm}(\bm{\theta}\hspace{-0.04cm}-\hspace{-0.03cm}\bm{\theta}_0))&\geq \sigma_{\textnormal{min}}(\mathbf{J}(\bm{\theta}_0)) \hspace{-0.06cm}-\hspace{-0.045cm} \|\mathcal{I}_{\bm{\theta}_0}\hspace{-0.02cm}(\bm{\theta}\hspace{-0.05cm}-\hspace{-0.04cm}\bm{\theta}_0)\hspace{-0.03cm}-\hspace{-0.03cm}\mathbf{J}(\bm{\theta}_0) \| \notag \\
    &\geq \beta \sigma_{\textnormal{min}}(\mathbf{J}(\bm{\theta}_0)), \notag 
\end{align}
which implies
\begin{align}
    \|\bm{\mathcal{G}}(\hspace{-0.01cm}\bm{\theta}\hspace{-0.01cm})\hspace{-0.06cm}-\hspace{-0.045cm}\bm{\mathcal{G}}(\hspace{-0.01cm}\bm{\theta}_0\hspace{-0.01cm})\|_{\textnormal{F}}&\geq \sigma_{\textnormal{min}}(\mathcal{I}_{\bm{\theta}_0}(\bm{\theta}-\bm{\theta}_0)) \|\bm{\theta}-\bm{\theta}_0\| \notag \\
    &\geq \beta\sigma_{\textnormal{min}}(\mathbf{J}(\bm{\theta}_0)) \|\bm{\theta}-\bm{\theta}_0\|, \notag 
\end{align}
which proves the statement. \hfill $\square$

\section{Computation of the asymp. variance of $\hat{a}_1$ and $\hat{b}_0$}
\label{sec:standardascov}
In this appendix we provide the computations that lead to the variance in \eqref{approxnormal} through an asymptotic approach. It is well known (see, e.g., \cite{ljung1998system}) that the covariance of $\hat{\bm{\theta}}$ can be approximated for large sample size by
\begin{equation}
\textnormal{Cov}\hspace{-0.02cm}\{ \hspace{-0.01cm}\hat{\bm{\theta}} \hspace{-0.01cm}\} \hspace{-0.09cm}\approx \hspace{-0.06cm}\frac{\sigma^2}{N} \hspace{-0.1cm}\left[\hspace{-0.02cm}\overline{\mathbb{E}}\hspace{-0.09cm}\left\{\hspace{-0.07cm}
\left(\hspace{-0.05cm}\dfrac{\partial \hat{y}(kh,\hspace{-0.02cm}\bm{\theta})}{\partial \bm{\theta}}\hspace{-0.02cm}\Big|_{\hspace{-0.01cm}\bm{\theta}=\bm{\theta}_0}\hspace{-0.06cm}\right)\hspace{-0.11cm}
\left(\hspace{-0.05cm}\dfrac{\partial \hat{y}(kh,\hspace{-0.02cm}\bm{\theta})}{\partial \bm{\theta}}\hspace{-0.02cm}\Big|_{\hspace{-0.01cm}\bm{\theta}=\bm{\theta}_0}\hspace{-0.06cm}\right)^{\hspace{-0.13cm}\top} \hspace{-0.04cm}\right\}\hspace{-0.04cm}\right]^{\hspace{-0.05cm}-\hspace{-0.02cm}1}\hspace{-0.17cm}, \notag
\end{equation}
where $\overline{\mathbb{E}}\{\cdot\}$ denotes the expectation of a quasi-stationary signal \cite[p. 34]{ljung1998system}, and the predictor $\hat{y}(kh,\bm{\theta})$ is given by
\begin{equation}
\hat{y}(kh,\bm{\theta}) = \alpha_1 |G(\mathrm{i}\omega_1,\bm{\theta})| \sin\big(\omega_1 kh + \angle G(\mathrm{i}\omega_1,\bm{\theta})\big). \notag
\end{equation}
After standard computations, we obtain
\begin{equation}
\dfrac{\partial \hat{y}(kh,\bm{\theta})}{\partial \bm{\theta}}\Big|_{\bm{\theta}=\bm{\theta}_0} = \begin{bmatrix}
\dfrac{-\alpha_1 b_{0,0} \omega_1}{a_{1,0}^2\omega_1^2 + 1}\cos(\omega_1 kh+2\phi) \\
\dfrac{\alpha_1}{\sqrt{a_{1,0}^2\omega_1^2 + 1}}\sin(\omega_1 kh+\phi)
\end{bmatrix}, \notag 
\end{equation}
where $\phi=\angle (1-\mathrm{i} a_{1,0}\omega_1)$. Using the identity
\begin{equation}
\overline{\mathbb{E}}\left\{\sin(\omega_1 kh + \phi_1)\sin(\omega_1 kh + \phi_2)\right\} = \frac{\cos(\phi_1-\phi_2)}{2}, \notag
\end{equation}
we obtain
\begin{align}
\overline{\mathbb{E}}&\left\{
\left(\dfrac{\partial \hat{y}(kh,\bm{\theta})}{\partial \bm{\theta}}\Big|_{\bm{\theta}=\bm{\theta}_0}\right)
\left(\dfrac{\partial \hat{y}(kh,\bm{\theta})}{\partial \bm{\theta}}\Big|_{\bm{\theta}=\bm{\theta}_0}\right)^\top \right\}^{-1} \notag \\
&\hspace{-0.1cm}= \hspace{-0.05cm}\frac{4(a_{1,0}^2\omega_1^2+1)^4}{\alpha_1^4 b_{0,0}^2 \omega_1^2}\hspace{-0.05cm}\begin{bmatrix}
\frac{\alpha_1^2}{2(a_{1,0}^2\omega_1^2+1)} & \hspace{-0.2cm} \frac{-\alpha_1^2 b_{0,0} \omega_1\sin(\phi)}{2(a_{1,0}^2\omega_1^2+1)^{3/2}} \\
\frac{-\alpha_1^2 b_{0,0} \omega_1\sin(\phi)}{2(a_{1,0}^2\omega_1^2+1)^{3/2}} & \hspace{-0.2cm}\frac{\alpha_1^2 b_{0,0}^2 \omega_1^2}{2(a_{1,0}^2\omega_1^2+1)^2}
\end{bmatrix}. \notag
\end{align}
Thus, we conclude that
\begin{equation}
\textnormal{Var}\{\hspace{-0.01cm}\hat{a}_1\hspace{-0.01cm}\} \hspace{-0.06cm}\approx\hspace{-0.05cm} \frac{2\sigma^2\hspace{-0.03cm}(a_{1,0}^2\omega_1^2\hspace{-0.04cm}+\hspace{-0.05cm}1)^3}{N\alpha_1^2 b_{0,0}^2\omega_1^2},\hspace{0.1cm} \textnormal{Var}\{\hspace{-0.01cm}\hat{b}_0\hspace{-0.01cm}\} \hspace{-0.06cm}\approx \hspace{-0.04cm}\frac{2\sigma^2(a_{1,0}^2\omega_1^2\hspace{-0.04cm}+\hspace{-0.05cm}1)^2}{N\alpha_1^2}. \notag
\end{equation}

\section*{References}
\vspace{-0.6cm}
\bibliographystyle{IEEEtran}
\balance
\bibliography{References}

\end{document}